\documentclass[%
superscriptaddress,
reprint,
preprintnumbers,
amsmath,amssymb,aps,prl]{revtex4-1}

\usepackage{bm}
\usepackage[colorlinks=true,linktocpage=true,linkcolor=blue,citecolor=blue,urlcolor=blue]{hyperref}

\usepackage{cleveref} 

\usepackage[utf8]{inputenc} 
\usepackage[T1]{fontenc} 
\usepackage{microtype}

\usepackage[mathscr]{euscript}
\usepackage{relsize}

\usepackage{tensor}  

\usepackage{color,soul}

\newcommand{\half}{\frac{1}{2}}
\newcommand{\nn}{\nonumber}

\newcommand{\bfrac}[2]{\lb\frac{#1}{#2}\rb}

\newcommand{\lie}{\pounds}
\newcommand{\df}{\mathrm{d}}
\newcommand{\dow}{\partial} 

\usepackage{mathtools}

\usepackage{xparse}

\ExplSyntaxOn

\clist_map_inline:nn{A,B,C,D,E,F,G,H,I,J,K,L,M,N,O,P,Q,R,S,T,U,V,W,X,Y,Z}
{
  \expandafter\def\csname b#1\endcsname{\bm{#1}} 
  
  \expandafter\def\csname c#1\endcsname{\mathcal{#1}} 
  
  \expandafter\def\csname bc#1\endcsname{\bm{\mathcal{#1}}} 
  
  \expandafter\def\csname s#1\endcsname{{\mathsmaller{#1}}} 
  
  \expandafter\def\csname bb#1\endcsname{\mathbb{#1}} 
  
  \expandafter\def\csname rm#1\endcsname{\mathrm{#1}} 
  
  \expandafter\def\csname sc#1\endcsname{\mathscr{#1}} 
  
  \expandafter\def\csname sf#1\endcsname{\mathsf{#1}} 
  
  \expandafter\def\csname f#1\endcsname{\mathfrak{#1}} 
}

\ExplSyntaxOff

\let\Im\relax
\DeclareMathOperator{\Im}{Im}
\let\Re\relax
\DeclareMathOperator{\Re}{Re}

\newcommand{\lB}{\left [}
\newcommand{\rB}{\right ]}
\newcommand{\lb}{\left (}
\newcommand{\rb}{\right )}

\newcommand\sbinom[2]{{\textstyle\binom{#1}{#2}}}

\usepackage{blindtext}
\usepackage[bbgreekl]{mathbbol}

\usepackage{tikz}
\usetikzlibrary{decorations.markings,decorations.pathmorphing,arrows.meta}

\tikzset{aux/.style={decorate,decoration={snake,segment length=1.5mm,
      amplitude=0.4mm}}}
\tikzset{left/.style={arrows={Stealth[scale length=0.5, scale width=1.5][sep=2pt]-}}}
\tikzset{right/.style={arrows={-[sep=-2pt]Stealth[scale length=0.5, scale
      width=1.5]}}}

\begin{document} 

\title{Non-universality of hydrodynamics}

\author{Akash Jain}\email{ajain@uvic.ca}
\author{Pavel Kovtun}\email{pkovtun@uvic.ca}

\affiliation{Department of Physics \& Astronomy, University of Victoria, PO
  Box 1700 STN CSC, Victoria, BC, V8W 2Y2, Canada.}


\begin{abstract} \noindent
  We investigate the effects of stochastic interactions on hydrodynamic
  correlation functions using the Schwinger-Keldysh effective field theory. We
  identify new ``stochastic transport coefficients'' 
  that are invisible in the classical constitutive relations, but
  nonetheless affect the late-time behaviour of hydrodynamic correlation functions
  through loop corrections. These results indicate that classical transport
  coefficients do not provide a universal characterisation of long-distance, late-time
  correlations even within the framework of fluctuating hydrodynamics.
\end{abstract}

\pacs{Valid PACS appear here}

\maketitle


Hydrodynamics is often referred to as the ``universal'' low-energy effective
description of many-body systems near thermal equilibrium. It is argued that if
one waits long enough for all the high-energy ``fast'' modes to thermalise, the
spectrum of a system can effectively be captured by the remaining ``slow'' modes
associated with conserved operators (such as energy, momentum, and particle
number). Fluctuations of conserved operators persist over long scales as they
need to be transported out to infinity to thermalise. A hydrodynamic system is
characterised by its fluxes expressed in terms of densities (or chemical
potentials) and their derivatives, known as \emph{constitutive relations}, with
dynamics governed by the associated conservation equations.

It is known that this ``classical'' picture of hydrodynamics is
incomplete. Hydrodynamic equations can be used to obtain the physically
observable retarded correlation functions; see~\cite{1963AnPhy..24..419K}. But
these results can potentially be contaminated by interactions between the slow
hydrodynamic modes and a background of fast
modes~\cite{Martin:1973zz,Pomeau:1974hg}. A more complete picture is offered by
the formalism of \emph{stochastic hydrodynamics}, wherein the collective
excitations of fast modes are modelled by random small-scale noise in the
hydrodynamic equations~\cite{Martin:1973zz, Pomeau:1974hg,
  Hohenberg:1977ym, DeDominicis:1977fw, Khalatnikov:1983ak}. The short-ranged
stochastic interactions are fine-tuned to reproduce the classical hydrodynamic
results at tree level. However, consistently including loop corrections one
finds that, for instance, the 2-point correlation function of fluid velocity has
non-analytic behaviour in $\omega$, referred to as ``long-time tails'', that is
not predicted by classical hydrodynamics~\cite{Forster:1976zz}.

This formalism, however, is not exhaustive as the requirement to reproduce
classical hydrodynamics does not uniquely fix the structure of stochastic
interactions. Importantly, assuming these random interactions to be Gaussian, as
is typically done, still leaves room for ambiguities. Physically, the ambiguities
correspond to some high-energy ``fast physics'', that has been ignored at the
classical level, leaking into the low-energy correlation functions via
interactions. This would mean that, contrary to what is typically believed, the
hydrodynamic transport coefficients \emph{do not} universally characterise the
low-energy spectrum of thermal systems.

The stochastic contamination in hydrodynamics can also be motivated from general
considerations in thermal field theory. Fluctuation-dissipation theorems (FDT)
imply that all the information in 2- and 3-point thermal correlation functions
in a system can be captured by the respective retarded functions. However, for
4- or higher-point correlations, retarded functions are no longer
enough~\cite{Wang:1998wg}. Classical hydrodynamics is only sensitive to
tree-level retarded correlations of conserved operators, and is consequently
blind to any information that might be contained in non-retarded higher-point
correlation functions. These higher-point correlations can nonetheless affect
the classically observable retarded functions of conserved densities and fluxes
through loop corrections. The point of this letter is to make the
above qualitative arguments precise and to explore the limits of hydrodynamics.

To probe these questions effectively, one needs a systematic prescription to
include stochastic noise into the hydrodynamic framework. While classically,
hydrodynamics is posed as a system of conservation equations, there now exists a
complete effective field theory (EFT) for hydrodynamics based on the
Schwinger-Keldysh (SK) formalism of thermal field
theories~\cite{Haehl:2015uoc, Crossley:2015evo,
  Jensen:2017kzi}; see~\cite{Glorioso:2018wxw} for a review. The effective
Kubo-Martin-Schwinger (KMS) condition in this formalism ensures that the FDT
requirements are built into the EFT, and can be used to investigate stochastic
signatures in hydrodynamic response functions. This formalism has recently been
used to revisit long-time tails due to diffusive fluctuations at one-loop order
in~\cite{Chen-Lin:2018kfl}.

We argue that the effective action for hydrodynamics can be naturally organised
into what we call ``KMS blocks.'' The first KMS block contains all the
information about fully retarded tree-level correlation functions, i.e. classical
hydrodynamics, plus a minimal set of higher-point interactions enforced by KMS
conditions. Aligning with the expectations from thermal field theory, the second
KMS block starts at the level of four-point interactions and contains all the
residual information about correlation functions that are retarded in all but
one momenta, and so on for higher KMS blocks. Interactions in the $n$th KMS
block are typically suppressed with at least $(2n-1)$ derivatives compared to
ideal hydrodynamics, therefore the first stochastic signatures creep into
hydrodynamics at third derivative order. This signals a non-universality of
higher-derivative corrections in hydrodynamics.

\vspace{1em}

\emph{Stochastic interactions in simple diffusion.}---For a concrete realisation
of these ideas, we consider a simplified model with a single conserved density
$J^t = n(\mu)$, where $\mu$ is the associated chemical potential. Classical
evolution of $n$ is governed by its conservation (diffusion) equation
$\dow_\mu J^\mu = 0$ with $J^i = -\sigma(\mu)\dow^i\mu$, with diffusion constant
given by $D = \sigma(\mu)/n'(\mu)$. The conductivity $\sigma(\mu)$ is a
non-negative classical ``transport coefficient''.

The EFT for diffusion is described by a phase field $\varphi_r$ and an
associated stochastic noise field $\varphi_a$~\cite{Crossley:2015evo}. We
introduce background gauge fields $\phi_{r,a} = (A_{r,a\mu})$ coupled to the
noise and physical current operators $\mathcal{O}_{a,r} = (J^\mu_{a,r})$
respectively. The effective action $S$ of the theory is constructed out of the
background gauge invariant building blocks
$\Phi_{r,a} = (B_{r,a\mu} = A_{r,a\mu}+\dow_\mu\varphi_{r,a})$. Connected
correlation functions of $\mathcal{O}_{r,a}$ are computed via a path integral
over the dynamical fields $\psi = (\varphi_{r,a})$, i.e. \footnote{We follow the
  conventions of~\cite{Crossley:2015evo} for correlation functions.}
\begin{align}
  G_{\alpha\ldots}
  = i^{n_a}\lb \frac{-i\delta}{\delta\phi_{\bar \alpha}} \ldots \rb
  \ln
  \int\mathcal{D}\psi\, \exp(i S),
  \label{eq:GreensFunctions}
\end{align}
where $\alpha=r,a$ and $\bar\alpha = a,r$ is its conjugate, while $n_a$ is the
number of $a$ type fields on the left. $G_{raa}$ computes the retarded function,
while $G_{rrr}$ computes the symmetric one, with all the remaining combinations
in between~\cite{Bellac:2011kqa}. The theory is required to satisfy a set of SK
constraints
\begin{subequations}
  \begin{gather}
    \hspace{-0.5em}
    S[\Phi_r,\Phi_a {=} 0;\beta]=0,~
    S[\Phi_r,-\Phi_a;\beta] = -S^*[\Phi_r,\Phi_a;\beta], \hspace{-0.6em}
    \label{eq:SK1} \\
    \Im S[\Phi_r,\Phi_a;\beta] \geq 0, 
    \label{eq:SK2} \\
    S[\Phi_r,\Phi_a;\beta]
    = S[\Theta\Phi_r,\Theta\Phi_a + i\Theta\lie_\beta\Phi_r; \Theta\beta].
    \label{eq:KMS}
  \end{gather}
  \label{eq:SKsymm}%
\end{subequations}
Here $\lie_\beta$ denotes a Lie derivative along the thermal vector
$\beta^\mu = 1/T_0\,\delta^\mu_t$, with $T_0$ being the constant global
temperature, and $\Theta = \text{PT}$ represents a discrete spacetime parity
transformation. In particular, \eqref{eq:SK2} implements the inequality in the
second law of thermodynamics, while the KMS symmetry \eqref{eq:KMS} implements
FDT \footnote{We have restricted to the KMS symmetry in the statistical (small
  $\hbar$) regime.}.  The theory also has a local phase shift symmetry
\begin{equation}
  \varphi_r(x) \to \varphi_r(x) - \lambda(x)
  ~~\text{such that}~~
  \beta^\mu \dow_\mu \lambda(x) = 0.
  \label{eq:fluid-spacetime-U1}
\end{equation}

Given these requirements, at leading order in derivatives, we find the effective
Lagrangian~\cite{Crossley:2015evo}
\begin{subequations}
  \begin{align}
    \mathcal{L}_{1}
    &= n(\mu) B_{at}
      + i T_0\sigma(\mu) B_{ai} \lb B^i_a + i \lie_\beta B^i_r \rb,
      \label{eq:diff-Lcl} 
  \end{align}
  Here $\mu = B_{rt} = \dow_t\varphi_r + A_{rt}$. Given that $\sigma$ is
  non-negative, conditions \eqref{eq:SK1} and \eqref{eq:SK2} are trivially
  satisfied. The second term maps to itself under \eqref{eq:KMS}, while the
  first term merely generates an additional total derivative term
  $\dow_t p(\mu)$ such that $p'(\mu) = n(\mu)$. The classical diffusion equation
  can be recovered upon varying the action with respect to $\varphi_a$,
  restricting to configurations with $\varphi_a = 0$, and setting the background
  to $A_{r\mu} = \mu_0 \delta^t_\mu$, $A_{a\mu}=0$.

  While the action \eqref{eq:diff-Lcl} is sufficient to reproduce classical
  evolution, the formalism does allow for extra terms that are at least
  quadratic in noise fields and hence leave the classical dynamics
  untouched. For instance
  \begin{align}
    \mathcal{L}_{2}
    &= iT_0^2\vartheta_1(\mu) B_{ai}B_{aj}
      \lb \lie_\beta B_r^i \lie_\beta B_{r}^j
      - \delta^{ij} \lie_\beta B_r^k \lie_\beta B_{rk} \rb \nn\\
    & \hspace{-1em}
      + iT_0^2\vartheta_2(\mu) B^i_{a} B_{ai}
      \lb B^j_a \,{+}\, i\lie_\beta B^j_r \rb 
      \lb B_{aj} \,{+}\, i\lie_\beta B_{rj} \rb ,
      \label{eq:diff-Lst}
  \end{align}
  \label{eq:diff-L}%
\end{subequations}%
where $\vartheta_{1,2}(\mu)$ are arbitrary ``stochastic coefficients''. Each
term here involves at least four fields, so the stochastic coefficients only
contribute to 4- and higher-point non-fully-retarded correlation function at
tree level, as argued in the introduction. For example, denoting ``$r$'' type
fields by solid and ``$a$'' type by wavy lines, the partially-retarded function
$G_{rraa}$ of $n$ receives a tree-level stochastic contribution due to
interactions in \eqref{eq:diff-Lst} (see appendix)
\begin{subequations}
\begin{equation}
  \raisebox{-20pt}{\tikz[thick]{
      \draw [right] (-7mm,7mm)--(-3.5mm,3.5mm);
      \draw [aux] (-3.5mm,3.5mm)--(0,0);
      \node at (-7mm,3mm) {$p_1$};
      \draw [right] (-7mm,-7mm)--(-3.5mm,-3.5mm);
      \draw [aux] (-3.5mm,-3.5mm)--(0,0);
      \node at (-7mm,-3mm) {$p_2$};
      \draw [aux] (7mm,7mm)--(3.5mm,3.5mm);
      \draw [right] (0mm,0)--(3.5mm,3.5mm);
      \node at (7.5mm,3mm) {$p_2$};
      \draw [aux] (7mm,-7mm)--(3.5mm,-3.5mm);
      \draw [right] (0mm,0)--(3.5mm,-3.5mm);
      \node at (7.5mm,-3mm) {$p_1$};
      \draw[fill] (0,0) circle (.4ex);
    }}\quad
  \begin{aligned}
    G_{rraa}
    &= \ldots + \frac{2\omega^2 k^4}{(\omega+iDk^2)^4} \times \\
    & \quad \lb 2\vartheta_2 \cos^2\theta - \vartheta_1 \sin^2\theta \rb.
  \end{aligned}
  \label{eq:kubo}
\end{equation}
for $p_1 = (\omega,k,0,0)$ and $p_2 = (\omega,k\cos\theta,k\sin\theta,0)$. Here
$D = \sigma/\chi$ is the diffusion constant and $\chi = \dow n/\dow\mu$ is the
susceptibility. Ellipsis denote further non-stochastic corrections due to terms
in \cref{eq:diff-Lcl}. One can use the retarded functions $G_{raaa}$, $G_{raa}$,
$G_{ra}$ to cancel these terms, and obtain a Kubo formula for $\vartheta_1$,
$\vartheta_2$ using \eqref{eq:kubo}.

Although stochastic coefficients do not contribute to fully-retarded correlation
functions at tree-level, they do show up in the loop corrections such as
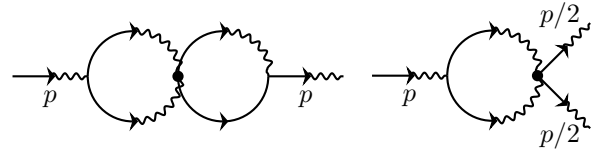
\begin{equation*}
  \raisebox{-10pt}{\tikz[thick]{
      \draw [right] (-10mm,0)--(-5mm,0);
      \draw [aux] (-5mm,0)--(0,0);
      \node at (-5mm,-3mm) {$p$};
      \draw [right] (0,0) arc (180:90:6mm);
      \draw [aux] (6mm,6mm) arc (90:0:6mm);
      %
      \draw [right] (0,0) arc (180:270:6mm);
      \draw [aux] (6mm,-6mm) arc (-90:0:6mm);
      %
      \draw[fill] (12mm,0) circle (.4ex);
      \draw [right] (12mm,0) arc (180:90:6mm);
      \draw [aux] (18mm,6mm) arc (90:0:6mm);
      %
      \draw [right] (12mm,0) arc (180:270:6mm);
      \draw (18mm,-6mm) arc (-90:0:6mm);
      %
      \draw [right] (24mm,0)--(29mm,0);
      \draw [aux] (29mm,0)--(34mm,0);
      \node at (29mm,-3mm) {$p$};
    }} \quad
  \raisebox{-20pt}{\tikz[thick]{
    \draw [right] (-10mm,0)--(-5mm,0);
    \draw [aux] (-5mm,0)--(0,0);
    \node at (-5mm,-3mm) {$p$};
    \draw [right] (0,0) arc (180:90:6mm);
    \draw [aux] (6mm,6mm) arc (90:0:6mm);
    %
    \draw [right] (0,0) arc (180:270:6mm);
    \draw [aux] (6mm,-6mm) arc (-90:0:6mm);
    \draw[fill] (12mm,0) circle (.4ex);
    \draw [right] (12mm,0)--(15.5mm,3.5mm);
    \draw [aux] (15.5mm,3.5mm)--(19mm,7mm);
    \node at (15mm,8mm) {$p/2$};
    \draw [right] (12mm,0)--(15.5mm,-3.5mm);
    \draw [aux] (15.5mm,-3.5mm)--(19mm,-7mm);
    \node at (15mm,-8mm) {$p/2$};
  }}
\end{equation*}
Here $p=(\omega,k,0,0)$. Stochastic vertices from \eqref{eq:diff-Lst} are
denoted in bold. We find that the retarded 2-point function of $n$ behaves in
$k^2 \ll \omega/D$ limit as (see appendix)
  \begin{align}
    \frac{\omega}{k^2} \Im G_{ra}
    &= \chi D
      + \frac{\chi^2\lambda^2 T_0}{32\pi D^{3/2}} \omega^{1/2}k^2
      + \ldots \nn\\
    &\qquad
      - \frac{\lambda^2 T_0 (\frac23\vartheta_1+\vartheta_2)}
      {1024\pi^2 D^4} \omega^2 k^4
      + \ldots.
      \label{eq:intro-G2}
  \end{align}
  Here $\lambda = \dow D/\dow n$. The leading correction to $G_{ra}$ dictated by
  the constitutive relations goes as $\omega^{1/2}k^2$~\cite{Chen-Lin:2018kfl},
  while the leading correction due to stochastic coefficients goes as
  $\omega^2k^4$. The retarded 3-point function, on the other hand, receives a
  non-analytic stochastic correction
  \begin{equation}
    -\frac{2\omega^2}{k^4} \Re G_{raa}
    = \chi^2 D \tilde{\lambda}
    + \ldots
    + \frac{\lambda(\frac23\vartheta_1+\vartheta_2)}{32 \pi
      D^{5/2}} \omega^{5/2} + \ldots.
    \label{eq:intro-G3}
  \end{equation}%
  \label{eq:intro-results}%
\end{subequations}
Here $\tilde\lambda = \dow D/\dow n + D/\chi\, \dow\chi/\dow n$. The middle
ellipsis in \eqref{eq:intro-results} denote subleading terms coming from
\eqref{eq:diff-Lcl}, while the final ellipsis denote terms higher order in
$k^2$. Detailed calculations for finite $k^2$ are given in the appendix. These
results illustrate that the hydrodynamic correlation functions start to receive
higher-derivative corrections that are \emph{not fixed by the constitutive
  relations}.

We note that the $\vartheta_1$-contribution to the effective action
\eqref{eq:diff-Lst} is quadratic in the noise field. Gaussian noise can be
captured by the conventional stochastic model, wherein one introduces a random
microscopic term $r^i$ in the flux $J^i = -\sigma(\mu)\dow^i\mu +
r^i$. Correlation functions are obtained by path integrating over $r^i$ weighted
by a Gaussian factor
$\exp(-1/4 \int \df^4 x\, r^i r^j
\lambda_{ij}[\mu])$~\cite{Kovtun:2012rj}. Imposing FDT constrains the form of
the coefficient $\lambda_{ij}$ in terms of hydrodynamic transport
coefficients. At leading order in derivatives, FDT uniquely fixes
$\lambda_{ij} = \delta_{ij}/(T_0\sigma(\mu))$. However, this uniqueness is
violated by higher derivative corrections pertaining to stochastic coefficients,
such as $\vartheta_{1,2}$, that are \emph{not} fixed by FDT. For example, the
$\vartheta_1$-term from \eqref{eq:diff-Lst} appears as
$\lambda_{ij} \sim -\vartheta_1(\mu)/(T_0\sigma(\mu))^2(\dow_i\mu\dow_j\mu -
\delta_{ij}\dow_k\mu\dow^k\mu)$.

\vspace{1em}

\emph{Stochastic interactions in hydrodynamics.}---The EFT for full relativistic
hydrodynamics is considerably more involved, but the discussion of stochastic
interactions follows a similar flow. In addition to the phase pair $\varphi_r$,
$\varphi_a$ associated with density fluctuations, the theory also contains the
Lagrangian coordinates $\sigma^{\sA=0,\ldots, 3}$ of the fluid elements and
respective noise $X_a^\mu$ as fundamental fields associated with energy-momentum
fluctuations~\cite{Crossley:2015evo}~\footnote{This is the so-called ``physical
  spacetime'' formulation of hydrodynamic EFT. See~\cite{Glorioso:2018wxw} for
  more details.}. We take $\sigma^0$ to define the local rest frame associated
with the global thermal state. The thermal vector $\beta^\mu$ is no longer a
constant, but is given by
$\beta^\mu = 1/T_0\, \dow x^\mu(\sigma(x))/\dow \sigma^0(x)$. Introducing
background fields $\phi_{r,a} = (g_{r,a\mu\nu},A_{r,a\mu})$ coupled to noise and
physical energy-momentum tensor/charge current operators
$\mathcal{O}_{a,r} = (T^{\mu\nu}_{a,r},J^\mu_{a,r})$ respectively, the
correlation functions can be computed by \eqref{eq:GreensFunctions}, with the
path integral over $\psi = (\varphi_{r,a},\sigma^\sA,X_a^\mu)$. The building
blocks for the respective effective action $S$, besides $\beta^\mu$, are
(see~\cite{Glorioso:2018wxw})
\begin{gather}
  B_{r\mu} = A_{r\mu} + \dow_\mu\varphi_r,\quad
  B_{a\mu} = A_{a\mu} + \dow_\mu\varphi_a + \lie_{X_a}A_{r\mu}, \nn\\
  G_{r\mu\nu} = g_{r\mu\nu}, \quad
  G_{a\mu\nu} = g_{a\mu\nu} + \lie_{X_a} g_{r\mu\nu}.
\end{gather}
Denoting $\Phi_{r,a} = (G_{r,a\mu\nu},B_{r,a\mu})$, the SK constraints and phase
shift symmetry are still given by \eqref{eq:SKsymm},
\eqref{eq:fluid-spacetime-U1}. 

Expressing $S = \int\df^4x\sqrt{-g_r}\,\mathcal{L}$, up to leading order in
derivatives, the effective action for relativistic hydrodynamics satisfying
these requirements is given as
\begin{subequations}
  \begin{align}
    \mathcal{L}_1
    &= \half \lb \epsilon\,u^\mu u^\nu
      + p\,\Delta^{\mu\nu} \rb G_{a\mu\nu}
      + n\, u^\mu B_{a\mu} \nn\\
    &\hspace{-1.5em}
      + \frac{iT}{4}\lb 2\eta\Delta^{\mu\rho}\Delta^{\nu\sigma}
      {+} (\zeta{-}{\textstyle\frac{2}{d}}\eta) \Delta^{\mu\nu}\Delta^{\rho\sigma}\rb
      G_{a\mu\nu} \lb G_{a\mu\nu}{+}i\lie_\beta G_{r\mu\nu} \rb \nn\\
    & + iT\sigma\, \Delta^{\mu\nu} B_{a\mu}
      \lb B_{a\nu} + i\lie_{\beta}B_{r\nu} \rb,
      \label{eq:hydro-Lcl}
  \end{align}
  where $\Delta^{\mu\nu} = g_r^{\mu\nu} + u^\mu u^\nu$.  Velocity $u^\mu$ (with
  $u^\mu u_\mu = -1$), temperature $T$, and chemical potential $\mu$ are defined
  via $u^\mu/T = \beta^\mu$, $\mu/T = \beta^\mu B_{r\mu}$. Energy density
  $\epsilon$, pressure $p$, number density $n$, viscosities $\eta$, $\zeta$, and
  conductivity $\sigma$ are functions of $T$, $\mu$. They satisfy
  $\df p = s\df T + n\df\mu$, $\epsilon+p=Ts+\mu n$ for entropy density
  $s$. Condition \eqref{eq:SK2} requires $\eta$, $\zeta$, $\sigma$ to be
  non-negative. Classical conservation equations of hydrodynamics are obtained
  by varying the action with respect to $X^\mu_a$, $\varphi_a$ in a
  configuration with $X^\mu_a = \varphi_a = 0$, and setting the background to
  $g_{r\mu\nu} = \eta_{\mu\nu}$, $A_{r\mu} = \mu_0 \delta^t_\mu$,
  $g_{a\mu\nu} = A_{a\mu} = 0$.

  Similar to \eqref{eq:diff-Lst}, the full hydrodynamic action can also be
  modified with arbitrary stochastic terms based on the symmetries of the
  theory. For instance we have
  \begin{align}
    \mathcal{L}_{2}
    &= iT^2\vartheta_1
      \lb \Delta^{\mu\rho}\Delta^{\nu\sigma} -
      \Delta^{\mu\nu}\Delta^{\rho\sigma}\rb
      \lie_\beta B_{r\rho} \lie_\beta B_{r\sigma}\,
      B_{a\mu}B_{a\nu} \nn\\
    & \hspace{-1.5em}
      + iT^2\vartheta_2 \Delta^{\mu\nu} \Delta^{\rho\sigma}
      B_{a\mu} B_{a\nu}
      \lb B_{a\rho} \,{+}\, i\lie_\beta B_{r\rho} \rb 
      \lb B_{a\sigma} \,{+}\, i\lie_\beta B_{r\sigma} \rb  \nn\\
    & \hspace{-1.5em}
      + iT^2\vartheta_3 \lb
      \Delta^{\mu\mu'}\Delta^{\nu\nu'}\Delta^{\rho\rho'}\Delta^{\sigma\sigma'}
      - \Delta^{\mu\rho}\Delta^{\nu\sigma}\Delta^{\mu'\rho'}\Delta^{\nu'\sigma'}
      \rb
      \nn\\
    & G_{a\mu\nu}G_{a\rho\sigma}\, \lie_\beta G_{r\mu'\nu'} \lie_\beta G_{r\rho'\sigma'} \nn\\
    & \hspace{-1.5em}
      + iT^2\vartheta_4
      \Delta^{\mu\nu}\Delta^{\mu'\nu'}\Delta^{\rho\sigma}\Delta^{\rho'\sigma'}
      G_{a\mu\nu}G_{a\mu'\nu'} \nn\\
    & \lb G_{a\rho\sigma}+i\lie_\beta G_{r\rho\sigma} \rb
      \lb G_{a\rho'\sigma'}+i\lie_\beta G_{r\rho'\sigma'} \rb
      + \ldots,
      \label{eq:hydro-Lst}
  \end{align}%
\end{subequations}%
with $\vartheta_i$ being a few stochastic coefficients; we do not perform the
exhaustive counting exercise here.

Contributions from stochastic interactions in \eqref{eq:hydro-Lst} to
hydrodynamic response functions can be computed similar to
\eqref{eq:intro-results}. We leave this analysis for future work. We note,
however, that non-stochastic interactions in the simplified diffusion model only
set in at one-derivative order as opposed to full non-linear hydrodynamics where
momentum/velocity fluctuations in \eqref{eq:hydro-Lcl} lead to ideal-order
interactions; see~\cite{Kovtun:2012rj}. Since part of the derivative suppression
of stochastic signatures in \eqref{eq:intro-G2}, \eqref{eq:intro-G3} arises from
non-stochastic vertices, we expect this suppression to be relaxed in full
hydrodynamics.

The stochastic coefficients $\vartheta_i$ also arise in the context of
non-relativistic (Galilean) hydrodynamics, in complete analogy with its
relativistic counterpart.  The quantitative details can be worked out along the
lines of~\cite{Jain:2020vgc}.


\vspace{1em}

\emph{KMS blocks.}---In our discussion so far, we introduced stochastic terms in
the effective action by hand. In this section, we discuss a general procedure to
construct such terms and, in doing so, classify the generic structure of
stochastic interactions admissible by the hydrodynamic EFT. For a generic
thermal system, the effective Lagrangian can be organised as
\begin{equation}
  \mathcal{L} = \sum_{n=1}^\infty \mathcal{L}_n,
  \label{eq:KMS-Block-decomposition}
\end{equation}
where the $n$th ``KMS block'' $\mathcal{L}_n$ contains the most generic terms
involving $n$ number of ``$a$'' fields allowed by symmetries, plus a set of
terms with higher number of ``$a$'' fields required in order to satisfy KMS/FDT
requirements. By definition, classical dynamics of the system, and tree-level fully
retarded correlation functions $G_{ra\ldots a}$, are completely characterised by
$\mathcal{L}_1$.  Higher KMS blocks $\mathcal{L}_n$ for $n>1$ contain stochastic
interactions that contribute to tree-level non-fully-retarded correlators
$G_{r\ldots ra\ldots a}$ involving at least $n$ instances of ``$r$'' type
operators. The decomposition \eqref{eq:KMS-Block-decomposition} is not unique;
we can always redefine a KMS block with terms from higher KMS blocks. Such
ambiguity in $\mathcal{L}_1$ is precisely the non-universality of classical
hydrodynamics. Nonetheless, we provide a particular characterisation of KMS
blocks for the hydrodynamic effective Lagrangian satisfying the requirements
\eqref{eq:SKsymm}.

Condition~\eqref{eq:SK1} implies that the $\mathcal{L}$ can be arranged in a
power series in $\Phi_a$ starting from the linear term. We start with a
parametrisation (see appendix)
\begin{align}
  \mathcal{L}
  &= \cD_1(\Phi_a)
    + i \sum_{n=1}^\infty
    \cD_{2n}(\underbrace{\Phi_a,\ldots}_{\times n},
    \underbrace{\Phi_a {+} i \lie_\beta\Phi_r,
    \ldots}_{\times n}) \nn\\
  &\hspace{-0.5em}
    + \sum_{n=1}^\infty
    \cD_{2n+1}(\Phi_a{+}{\textstyle\frac{i}{2}}\lie_\beta\Phi_r,
    \underbrace{\Phi_a,\ldots}_{\times n},
    \underbrace{\Phi_a {+} i \lie_\beta\Phi_r,\ldots}_{\times n}).
    \label{eq:L-G}
\end{align}
Here $\cD_m$ are a set of totally-symmetric real multi-linear maps, allowing $m$
arguments, made out of $\Phi_r$ and $\beta^\mu$. Here $\times n$ denotes $n$
identical arguments. For instance, the diffusive Lagrangian \eqref{eq:diff-L}
corresponds to the choice
\begin{align}
  \mathcal{D}_1(X)
  &= n(\mu) X_t, \nn\\
  \mathcal{D}_2(X,Y)
  &= T_0\sigma(\mu) X_i Y^i
    + T_0^2\lb \vartheta_1(\mu) + {\textstyle\frac{2}{3}}\vartheta_2(\mu)\rb  \nn\\
  &\qquad
    (\lie_\beta
B^i_r \lie_\beta B^j_r - \delta^{ij}\lie_\beta B^k_r\lie_\beta B_{rk}) X_iY_j,
    \nn\\
  \mathcal{D}_4(W,X,Y,Z)
  &= T_0^2\vartheta_2(\mu)\,\delta^{(ij}\delta^{kl)}\, W_iX_jY_kZ_l,
    \label{eq:diff-maps}
\end{align}
for arbitrary vectors $W_{\mu}$, $X_{\mu}$, $Y_{\mu}$, $Z_{\mu}$. Recall that
$\mu = B_{rt}$ and $\Phi_{r,a}=(B_{r,a\mu})$. This form is particularly
convenient because each term in the summation is closed under KMS: the $n$
instances of $\Phi_a$, $\Phi_a+i\lie_\beta\Phi_r$ map to each other up to
$\Theta$, and $\Phi_a+i/2\,\lie_\beta\Phi_r$ maps to itself. Requiring
\eqref{eq:L-G} to respect \eqref{eq:SK2} and \eqref{eq:KMS} (up to a total
derivative), we find
\begin{gather}
  \cD_1(\lie_\beta\Phi_r) = \nabla_\mu\mathcal{N}_0^\mu, \qquad
  \cD_m(\Phi_a,\cdots)~\text{are $\Theta$-even}, \nn\\
  \cD_{2}(\Phi_a,\ldots)\big|_{\text{0-derivative}} \geq 0,
  \label{eq:KMS.condition.1}
\end{gather}
for some vector $\mathcal{N}_0^\mu$. Note that changing any argument of
$\mathcal{D}_m$ from $\Phi_a$ to $\lie_\beta\Phi_r$ flips its $\Theta$-parity,
therefore its contribution to $\mathcal{L}$ is generically \emph{not}
$\Theta$-even. For \eqref{eq:diff-maps}, these are satisfied with
$\mathcal{N}_0^\mu = p(\mu)\beta^\mu$ (such that $p'(\mu) = n(\mu)$) and
$\sigma(\mu)\geq0$.  Note that only $\mathcal{D}_{1,2,3}$ from \eqref{eq:L-G} can
contribute to the classical constitutive relations. These generically satisfy an
emergent second law of thermodynamics
\begin{equation}
  \nabla_\mu S^\mu = \cD_2(\lie_\beta\Phi_r,\lie_\beta\Phi_r)
  + \frac{1}{2} \mathcal{D}_3(\lie_\beta\Phi_r,
  \lie_\beta\Phi_r,\lie_\beta\Phi_r),
  \label{eq:EC}
\end{equation}
for some $S^\mu$; see appendix. \eqref{eq:KMS.condition.1} guarantees the
positivity of entropy production, within the derivative expansion.

Generically, $\cD_n$ contain all structures allowed by symmetries at a given
derivative order. We refer to the contribution of each such structure in the
effective action \eqref{eq:L-G} as a ``KMS group''. Each KMS group is
independently invariant under the KMS transformation. The ``$n$th KMS block''
can be defined as the set of all KMS groups wherein the least nonzero power of
$\Phi_a$ fields is $n$. Inspecting \eqref{eq:L-G}, it follows that each group in
$\cD_n$ falls at least in the $\lfloor n/2\rfloor$-th KMS block. Here
$\lceil n/2\rceil$ and $\lfloor n/2\rfloor$ denote ceiling and floor
functions. We say ``at least'' because there can be groups in $\cD_n$ that
identically vanish (up to a total derivative) when one or more of their
arguments are $\lie_\beta\Phi_r$, e.g. $\vartheta_{1,2}$ contribution in
$\mathcal{D}_2$ in \eqref{eq:diff-maps}. As seen from \eqref{eq:L-G}, such
groups are pushed to higher KMS blocks.  To account for these subtleties, we can
decompose $\cD_{n} = \sum_{m=0}^{n} \cD_{n,m}$, where $\cD_{n,m}$ can be thought
of as the component of $\cD_{n}$ with $m$ of its arguments projected transverse
to $\lie_\beta \Phi_r$. More precisely, it contains groups from $\mathcal{D}_n$
that do not vanish for up to $n-m$ instances of $\lie_\beta\Phi_r$ in their
arguments, but vanish for $n-m+1$ instances. In particular, $\cD_{n,0}$ does not
vanish for any number of $\lie_\beta\Phi_r$, while $\cD_{n,n}$ vanishes for even
one. Note that $\cD_{1,0} = 0$ due to \eqref{eq:KMS.condition.1}. 

Plugging this decomposition into \eqref{eq:L-G}, we can work out the first KMS
block explicitly as
\begin{align}
  \mathcal{L}_1
  &= \mathcal{D}_{1,1}(\Phi_a)
    + i\mathcal{D}_{2,1}(\Phi_a,\Phi_a{+}i\lie_\beta\Phi_r) \nn\\
    &\quad
      + i\mathcal{D}_{2,0}(\Phi_a,\Phi_a{+}i\lie_\beta\Phi_r) \nn\\
  &\quad
    + \mathcal{D}_{3,0}(\Phi_a{+}{\textstyle\frac{i}{2}}\lie_\beta\Phi_r,
    \Phi_a,\Phi_a{+}i\lie_\beta\Phi_r),
    \label{eq:L1-general}
\end{align}
which completely characterises classical hydrodynamics. The first two terms
correspond to ``adiabatic'' transport, as they do not contribute to entropy
production in \eqref{eq:EC}. Their respective contribution to the constitutive
relations is $\Theta$-even and $\Theta$-odd respectively. The last two terms
correspond to $\Theta$-odd and $\Theta$-even ``dissipative'' transport leading
to entropy production in \eqref{eq:EC}. From our examples in \eqref{eq:diff-Lst}
and \eqref{eq:hydro-Lst}, $\epsilon,p,n\in\mathcal{D}_{1,1}$ and
$\eta,\zeta,\sigma\in\mathcal{D}_{2,0}$; other contributions show up at higher
order in derivatives. Technically, $\mathcal{D}_{3,1}$ also appears in
\eqref{eq:L1-general}, but can be pushed to $\mathcal{L}_2$ by redefining
$\mathcal{D}_{1,1}$; see appendix.


The first non-trivial stochastic corrections to classical hydrodynamics come
from the 2nd KMS block. Up to leading order in derivatives, this is given as
\begin{align}
  \mathcal{L}_2
  &= i\cD_{22}(\Phi_a,\Phi_a{+}i\lie_\beta\Phi_r)
    + \mathcal{D}_{3,1}(\Phi_a,\Phi_a,
    \Phi_a{+}{\textstyle\frac{3i}{2}}\lie_\beta\Phi_r)
    \nn\\
  &\hspace{-1em}
    + i\cD_{4,0}(\Phi_a,\Phi_a,\Phi_a{+}i\lie_\beta\Phi_r,\Phi_a{+}i\lie_\beta\Phi_r)
    +\mathcal{O}(\dow^5).
    \hspace{-0.1em}
    \label{eq:L2-general}
\end{align}
At this point, we are unable to ascertain any physical distinction between
various contributions. In our examples,
$(\vartheta_1+\frac{2}{3}\vartheta_2),\vartheta_3\in\mathcal{D}_{2,2}$, while
$\vartheta_2,\vartheta_4\in\cD_{4,0}$. Higher KMS blocks can be worked out in a
similar manner.

The count derivative ordering in hydrodynamics, we use the canonical scheme
from~\cite{Crossley:2015evo}, where $\Phi_r\sim\mathcal{O}(\dow^0)$ and
$\Phi_a,\lie_\beta\Phi_r\sim\mathcal{O}(\dow^1)$. Projecting $m$ arguments
against $\lie_\beta\Phi_r$ in $\mathcal{D}_{n,m}$ requires the introduction of
$m$ copies of $\lie_\beta\Phi_r$. Hence, the contribution of $\mathcal{D}_{n,m}$
to $\mathcal{L}$, and to the hydrodynamic observables, is typically suppressed
with $\mathcal{O}(\dow^{n+m-1})$ compared to the ideal order thermodynamic terms
in $\mathcal{D}_{1,1}$. Consequently, effects of stochastic KMS blocks
$\mathcal{L}_n$ for $n>1$ are suppressed with $\mathcal{O}(\dow^{2n-1})$
compared to $\mathcal{L}_1$, in addition to any loop-suppression.
 
\vspace{1em} 

\emph{Outlook.}---Hydrodynamics is an effective theory and an immensely
successful one at that. However, like any effective theory, it has a limited
scope of applicability. It posits that the low-energy dynamics of a many-body
thermal system can be effectively captured by the long-range transport
properties of its conserved charges. While it is certainly true to a leading
approximation, short-range stochastic interactions must be included into the
framework to consistently describe interactions between hydrodynamic modes. In
this letter, we took this fine-print a step further and investigated the
sensitivity of hydrodynamics to the choice of stochastic interactions.

We used the EFT framework of hydrodynamics developed
recently~\cite{Crossley:2015evo} and identified new ``stochastic transport
coefficients'' in \eqref{eq:diff-Lst} and \eqref{eq:hydro-Lst} characterising
short-range information. The stochastic coefficients do not enter the classical constitutive relations,
but nonetheless affect retarded correlation functions in the hydrodynamic regime at
subleading orders in derivatives through loop interactions. We explicitly
derived the stochastic signatures in 2- and 3-point retarded functions for
diffusive fluctuations in $(3+1)$ dimensions in \eqref{eq:intro-results}. In
particular, we found the stochastic correction to 3-point function to be
non-analytic in frequency at one-loop order. It is worth noting that these
results are different from the usual ``long-time tails'' as the effects we are describing
are characterised by entirely new transport coefficients which are invisible in classical
constitutive relations. Finally, we classified the general structure of
stochastic interactions through KMS blocks. Classical physics is completely
characterised by the first KMS block, while the higher KMS blocks characterise a
plethora of stochastic coefficients.

We conclude that the sensitivity of hydrodynamic observables to short-range
stochastic information signifies a breakdown of the celebrated universality of
hydrodynamics in describing long-range correlations.
It would be
interesting to find physical systems where the signatures of stochastic
coefficients are enhanced enough to overcome the loop suppression. As we
already discussed in the letter, part of this could be achieved by revisiting
the computation in the presence of momentum modes in Galilean or relativistic hydrodynamics. The stochastic signatures
are also enhanced in lower spatial dimensions. We plan to return to these 
questions in the future.

\vspace{1em}

This work was supported in part by the NSERC Discovery Grant program of Canada.

\bibliography{mySpires_ajain,nonhydro} 


\newpage

\appendix
\renewcommand\thefigure{A\arabic{figure}}    
\setcounter{figure}{0} 
\renewcommand{\theequation}{A\arabic{equation}}
\setcounter{equation}{0}

{\center{\large\bfseries Supplementary Material}\par}

\section{Linearised fluctuations in diffusion EFT}

In this appendix we give details of loop calculations in diffusion EFT. Analysis
for full hydrodynamics proceeds in a similar manner.

\vspace{1em}

\emph{Linearised action.}---To compute various correlation functions, we need to
expand the effective action order-by-order in self interactions. It is
convenient to work with the density $n$ as a fundamental degree of freedom
instead of $\mu$. We can expand the Lagrangian \eqref{eq:diff-Lcl} up to forth
order in the fields $\delta n = n - n(\mu_0)$ and $\varphi_a$, in the absence of
background fields, to obtain
\begin{align}
  \mathcal{L}^{\text{free}}_{1}
  &= - \varphi_a \lb \dow_t \delta n - D \dow^2 \delta n \rb
  + iT_0\sigma \dow^i\varphi_a\dow_i\varphi_a , \nn\\
  \mathcal{L}_{1}^{\text{3pt}}
  &= \half \lambda \delta n^2 \dow^2 \varphi_a
    + i\chi T_0 \tilde\lambda \delta n \dow^i\varphi_a\dow_i\varphi_a ,
    \nn\\
  \mathcal{L}_{1}^{\text{4pt}}
  &= \frac13 \lambda_4 \delta n^3 \dow^2 \varphi_a
    + i\chi T_0 \tilde\lambda_4 \delta n^2 \dow^i\varphi_a\dow_i\varphi_a.
    \label{eq:diff-Lcl-expanded}
\end{align}
Here $\chi = \dow n/\dow\mu$ is the susceptibility and $D = \sigma/\chi$ is the
diffusion constant, along with
\begin{gather}
  \lambda = \frac{1}{\chi} \frac{\dow D}{\dow\mu}, \quad \lambda_4 =
  \frac{1}{2\chi} \frac{\dow}{\dow\mu}
  \lb \frac{1}{\chi} \frac{\dow D}{\dow\mu}\rb, \nn\\
  \tilde\lambda = \frac{1}{\chi^2} \frac{\dow\sigma}{\dow\mu}, \qquad
  \tilde\lambda_4 = \frac{1}{2\chi^2} \frac{\dow}{\dow\mu} \lb \frac{1}{\chi}
  \frac{\dow\sigma}{\dow\mu}\rb.
\end{gather}
In a typical diffusive model, $\omega\sim k^2$. Taking $D,\chi,T_0\sim 1$, and
noting that $\mathcal{L} \sim k^d\omega \sim k^{d+2}$, we can infer that
$\varphi_a,\delta n\sim k^{d/2}$. Therefore, higher order interactions in
$\delta n$ and $\varphi_a$ are successively more irrelevant in $k$ and can be
consistently dropped within the derivative expansion. This form of the diffusive
action was recently derived in~\cite{Chen-Lin:2018kfl}. The coefficients
$\lambda_4$ and $\tilde\lambda_4$ are denoted as $\lambda'$ and $\tilde\lambda'$
in~\cite{Chen-Lin:2018kfl}; we reserve primes to denote derivatives with respect
to $\mu$. For the stochastic Lagrangian \eqref{eq:diff-Lst}, we get the first
non-trivial contribution as
\begin{align}
  \mathcal{L}_{2}^{\text{4pt}}
  &= i\frac{\vartheta_1}{\chi^2} (\dow^in \dow_i\varphi_a)^2
    -i\frac{\vartheta_1+\vartheta_2}{\chi^2} (\dow^i n \dow_i n)
    (\dow^j\varphi_a\dow_j\varphi_a) \nn\\
  &\hspace{-3em}
  - \frac{2T_0 \vartheta_2}{\chi} (\dow^i n \dow_i\varphi_a)
    (\dow^j\varphi_a\dow_j\varphi_a) 
    + iT_0^2 \vartheta_2 (\dow^i\varphi_a\dow_i\varphi_a)^2.
    \hspace{-0.5em}
    \label{eq:diff-Lst-expanded}
\end{align}

$\mathcal{L}^{\text{free}}_{1}$ in \eqref{eq:diff-Lcl-expanded} is the free
Lagrangian and leads to the tree propagators
\begin{align}
  \langle \delta n(p)\varphi_a(-p) \rangle_0
  &= \frac{1}{F(p)}
  &= \raisebox{-12pt}{\tikz[thick]{
    \draw [right] (-14mm,0)--(-7mm,0);
    \draw [aux] (-7mm,0)--(0,0);
    \node at (-7mm,-3mm) {$p$};
    }}
  \nn\\
  \langle \varphi_a(p)\delta n(-p) \rangle_0
  &= \frac{-1}{F(p)^*}
  &= \raisebox{-12pt}{\tikz[thick]{
    \draw [right,aux] (-14mm,0)--(-7mm,0);
    \draw (-7mm,0)--(0,0);
    \node at (-7mm,-3mm) {$p$};
    }} \nn\\
  \langle \delta n(p)\delta n(-p) \rangle_0
  &= \frac{2T_0\chi D k^2}{|F(p)|^2}
  &= \raisebox{-12pt}{\tikz[thick]{
    \draw [right] (-14mm,0)--(-7mm,0);
    \draw (-7mm,0)--(0,0);
    \node at (-7mm,-3mm) {$p$};
    }} \nn\\
  &= \frac{iT_0\chi}{F(p)}
  - \frac{iT_0\chi}{F(p)^*},
    \label{eq:diff-props}
\end{align}
where $p=(\omega,k)$ and $F(p) = \omega + iDk^2$. We denote $\delta n$ by solid
and $\varphi_a$ by wavy lines. The remaining terms in
\eqref{eq:diff-Lcl-expanded} and \eqref{eq:diff-Lst-expanded} lead to various
interaction vertices
\begin{gather}
  \raisebox{-29pt}{\tikz[thick]{
      \draw [left,aux] (-6mm,0)--(0,0);
      \node at (-5mm,-3mm) {$p_1$};
      \draw [right] (0mm,0)--(5mm,5mm);
      \node at (8mm,4mm) {$p_2$};
      \draw [right] (0mm,0)--(5mm,-5mm);
      \node at (8mm,-4mm) {$p_3$};
    }}~~
  \raisebox{-29pt}{\tikz[thick]{
      \draw [left,aux] (-6mm,0)--(0,0);
      \node at (-5mm,-3mm) {$p_1$};
      \draw [right,aux] (0mm,0)--(5mm,5mm);
      \node at (8mm,4mm) {$p_2$};
      \draw [right] (0mm,0)--(5mm,-5mm);
      \node at (8mm,-4mm) {$p_3$};
    }} \quad
  \raisebox{-29pt}{\tikz[thick]{
      \draw [left,aux] (-5mm,5mm)--(0,0);
      \node at (-7mm,4mm) {$p_1$};
      \draw [left] (-5mm,-5mm)--(0,0);
      \node at (-7mm,-4mm) {$p_2$};
      \draw [right] (0mm,0)--(5mm,5mm);
      \node at (8mm,4mm) {$p_3$};
      \draw [right] (0mm,0)--(5mm,-5mm);
      \node at (8mm,-4mm) {$p_4$};
    }}~~
  \raisebox{-29pt}{\tikz[thick]{
      \draw [left,aux] (-5mm,5mm)--(0,0);
      \node at (-7mm,4mm) {$p_1$};
      \draw [left,aux] (-5mm,-5mm)--(0,0);
      \node at (-7mm,-4mm) {$p_2$};
      \draw [right] (0mm,0)--(5mm,5mm);
      \node at (8mm,4mm) {$p_3$};
      \draw [right] (0mm,0)--(5mm,-5mm);
      \node at (8mm,-4mm) {$p_4$};
    }} \nn\\
  \raisebox{-15pt}{\tikz[thick]{
      \draw [left,aux] (-5mm,5mm)--(0,0);
      \node at (-7mm,4mm) {$p_1$};
      \draw [left,aux] (-5mm,-5mm)--(0,0);
      \node at (-7mm,-4mm) {$p_2$};
      \draw [right] (0mm,0)--(5mm,5mm);
      \node at (8mm,4mm) {$p_3$};
      \draw [right] (0mm,0)--(5mm,-5mm);
      \node at (8mm,-4mm) {$p_4$};
      \draw[fill] (0,0) circle (.4ex);
    }}~~
  \raisebox{-15pt}{\tikz[thick]{
      \draw [left,aux] (-5mm,5mm)--(0,0);
      \node at (-7mm,4mm) {$p_1$};
      \draw [left,aux] (-5mm,-5mm)--(0,0);
      \node at (-7mm,-4mm) {$p_2$};
      \draw [right,aux] (0mm,0)--(5mm,5mm);
      \node at (8mm,4mm) {$p_3$};
      \draw [right] (0mm,0)--(5mm,-5mm);
      \node at (8mm,-4mm) {$p_4$};
      \draw[fill] (0,0) circle (.4ex);
    }}~~
  \raisebox{-15pt}{\tikz[thick]{
      \draw [left,aux] (-5mm,5mm)--(0,0);
      \node at (-7mm,4mm) {$p_1$};
      \draw [left,aux] (-5mm,-5mm)--(0,0);
      \node at (-7mm,-4mm) {$p_2$};
      \draw [right,aux] (0mm,0)--(5mm,5mm);
      \node at (8mm,4mm) {$p_3$};
      \draw [right,aux] (0mm,0)--(5mm,-5mm);
      \node at (8mm,-4mm) {$p_4$};
      \draw[fill] (0,0) circle (.4ex);
    }}
\end{gather}
Stochastic vertices from \eqref{eq:diff-Lst-expanded} are denoted in bold. The
respective Feynman rules can be read off from directly from
\eqref{eq:diff-Lcl-expanded} and \eqref{eq:diff-Lst-expanded}. Energy-momentum
conservation at each vertex is understood.

One final piece of information that we need is the coupling to sources. As we
shall only be interested in correlation functions of density, we only keep the
scalar sources $A_{r,a\,t}$, and truncate to forth order in $A_{r,a\,t},\delta n,\varphi_a$,
\begin{align}
  \mathcal{L}_{1s}^{\text{2pt}}
  &= A_{at}\delta n
    + A_{rt}\chi \dow_t\varphi_a
    +  A_{at} A_{rt}\chi
    \nn\\
  \mathcal{L}_{1s}^{\text{3pt}}
  &= A_{rt}
    \lb
    \frac{\chi'}{\chi} \delta n\, \dow_t\varphi_a
    - \frac{\sigma'}{\chi} \dow^i\delta n \dow_i\varphi_a
    + i T_0 \sigma' \dow^i\varphi_a \dow_i\varphi_a \rb \nn\\
  &
    + A_{at} A_{rt} \frac{\chi'}{\chi} \delta n
    + \half A_{rt}^2 \chi' \dow_t\varphi_a
    + \half A_{at} A_{rt}^2 \chi' \nn\\
  \mathcal{L}_{1s}^{\text{4pt}}
  &= A_{rt} \lb \frac{1}{2\chi}\bfrac{\chi'}{\chi}' \delta n^2 \dow_t\varphi_a
    - \frac{\sigma''}{\chi^2}
    \delta n\, \dow^i\delta n\, \dow_i\varphi_a \rb \nn\\
  &
    + A_{rt} \frac{i T_0\sigma''}{\chi}
    \delta n \dow^i\varphi_a \dow_i\varphi_a
    + A_{at} A_{rt} \frac{1}{2\chi}\bfrac{\chi'}{\chi}'
    \delta n^2 \nn\\
  &+ \half A_{rt}^2  \lb \frac{\chi''}{\chi} \delta n \dow_t\varphi_a
    - \frac{\sigma'' }{\chi} \dow^i\delta n \dow_i\varphi_a
    + i T_0\sigma'' \dow^i\varphi_a\dow_i\varphi_a\rb
    \nn\\
  &     + \frac{\chi''}{2\chi} A_{at} A_{rt}^2 \delta n
    + \frac{\chi''}{6} \lb A_{at} + \dow_t\varphi_a \rb A_{rt}^3.
    \label{eq:diff-Ls-expanded}
\end{align}
We do not get any contribution from $\mathcal{L}_2$.  The first two terms in
$\mathcal{L}_{1s}^{\text{2pt}}$ are the usual linear couplings between
fundamental fields and sources, while the remaining non-linear couplings can be
represented by the vertices
\begin{gather}
  \raisebox{-29pt}{\tikz[thick]{
      \draw [left,aux] (-6mm,0)--(0,0);
      \node at (-5mm,-3mm) {$p_1$};
      \draw [right] (0mm,0)--(5mm,5mm);
      \node at (8mm,4mm) {$p_2$};
      \draw [right,dotted] (0mm,0)--(5mm,-5mm);
      \node at (8mm,-4mm) {$p_3$};
    }}~~
  \raisebox{-29pt}{\tikz[thick]{
      \draw [left,aux] (-6mm,0)--(0,0);
      \node at (-5mm,-3mm) {$p_1$};
      \draw [right,aux] (0mm,0)--(5mm,5mm);
      \node at (8mm,4mm) {$p_2$};
      \draw [right,dotted] (0mm,0)--(5mm,-5mm);
      \node at (8mm,-4mm) {$p_3$};
    }} \quad
    \raisebox{-29pt}{\tikz[thick]{
      \draw [left,dashed] (-6mm,0)--(0,0);
      \node at (-5mm,-3mm) {$p_1$};
      \draw [right] (0mm,0)--(5mm,5mm);
      \node at (8mm,4mm) {$p_2$};
      \draw [right,dotted] (0mm,0)--(5mm,-5mm);
      \node at (8mm,-4mm) {$p_3$};
    }}~~
  \raisebox{-29pt}{\tikz[thick]{
      \draw [left,aux] (-6mm,0)--(0,0);
      \node at (-5mm,-3mm) {$p_1$};
      \draw [right,dotted] (0mm,0)--(5mm,5mm);
      \node at (8mm,4mm) {$p_2$};
      \draw [right,dotted] (0mm,0)--(5mm,-5mm);
      \node at (8mm,-4mm) {$p_3$};
    }} \nn\\
  \raisebox{-29pt}{\tikz[thick]{
      \draw [left,aux] (-5mm,5mm)--(0,0);
      \node at (-7mm,4mm) {$p_1$};
      \draw [left] (-5mm,-5mm)--(0,0);
      \node at (-7mm,-4mm) {$p_2$};
      \draw [right] (0mm,0)--(5mm,5mm);
      \node at (8mm,4mm) {$p_3$};
      \draw [right,dotted] (0mm,0)--(5mm,-5mm);
      \node at (8mm,-4mm) {$p_4$};
    }}~~
  \raisebox{-29pt}{\tikz[thick]{
      \draw [left,aux] (-5mm,5mm)--(0,0);
      \node at (-7mm,4mm) {$p_1$};
      \draw [left,aux] (-5mm,-5mm)--(0,0);
      \node at (-7mm,-4mm) {$p_2$};
      \draw [right] (0mm,0)--(5mm,5mm);
      \node at (8mm,4mm) {$p_3$};
      \draw [right,dotted] (0mm,0)--(5mm,-5mm);
      \node at (8mm,-4mm) {$p_4$};
    }} \quad
  \raisebox{-29pt}{\tikz[thick]{
      \draw [left,dashed] (-5mm,5mm)--(0,0);
      \node at (-7mm,4mm) {$p_1$};
      \draw [left] (-5mm,-5mm)--(0,0);
      \node at (-7mm,-4mm) {$p_2$};
      \draw [right] (0mm,0)--(5mm,5mm);
      \node at (8mm,4mm) {$p_3$};
      \draw [right,dotted] (0mm,0)--(5mm,-5mm);
      \node at (8mm,-4mm) {$p_4$};
    }}~~
  \raisebox{-29pt}{\tikz[thick]{
      \draw [left,aux] (-5mm,5mm)--(0,0);
      \node at (-7mm,4mm) {$p_1$};
      \draw [left] (-5mm,-5mm)--(0,0);
      \node at (-7mm,-4mm) {$p_2$};
      \draw [right,dotted] (0mm,0)--(5mm,5mm);
      \node at (8mm,4mm) {$p_3$};
      \draw [right,dotted] (0mm,0)--(5mm,-5mm);
      \node at (8mm,-4mm) {$p_4$};
    }}~~\nn\\
  \raisebox{-29pt}{\tikz[thick]{
      \draw [left,aux] (-5mm,5mm)--(0,0);
      \node at (-7mm,4mm) {$p_1$};
      \draw [left,aux] (-5mm,-5mm)--(0,0);
      \node at (-7mm,-4mm) {$p_2$};
      \draw [right,dotted] (0mm,0)--(5mm,5mm);
      \node at (8mm,4mm) {$p_3$};
      \draw [right,dotted] (0mm,0)--(5mm,-5mm);
      \node at (8mm,-4mm) {$p_4$};
    }}\quad
  \raisebox{-29pt}{\tikz[thick]{
      \draw [left,dashed] (-5mm,5mm)--(0,0);
      \node at (-7mm,4mm) {$p_1$};
      \draw [left] (-5mm,-5mm)--(0,0);
      \node at (-7mm,-4mm) {$p_2$};
      \draw [right,dotted] (0mm,0)--(5mm,5mm);
      \node at (8mm,4mm) {$p_3$};
      \draw [right,dotted] (0mm,0)--(5mm,-5mm);
      \node at (8mm,-4mm) {$p_4$};
    }}~~
  \raisebox{-29pt}{\tikz[thick]{
      \draw [left,aux] (-5mm,5mm)--(0,0);
      \node at (-7mm,4mm) {$p_1$};
      \draw [left,dotted] (-5mm,-5mm)--(0,0);
      \node at (-7mm,-4mm) {$p_2$};
      \draw [right,dotted] (0mm,0)--(5mm,5mm);
      \node at (8mm,4mm) {$p_3$};
      \draw [right,dotted] (0mm,0)--(5mm,-5mm);
      \node at (8mm,-4mm) {$p_4$};
    }}~~\nn\\
  \raisebox{-12pt}{\tikz[thick]{
      \draw [dashed,left] (-6mm,0)--(0,0);
      \node at (-5mm,-3mm) {$p$};
      \draw [dotted,right] (0,0)--(6mm,0);
      \node at (5mm,-3mm) {$p$};
    }} \quad 
  \raisebox{-15pt}{\tikz[thick]{
      \draw [dashed,left] (-6mm,0)--(0,0);
      \node at (-5mm,-3mm) {$p_1$};
      \draw [right,dotted] (0mm,0)--(5mm,5mm);
      \node at (8mm,4mm) {$p_2$};
      \draw [right,dotted] (0mm,0)--(5mm,-5mm);
      \node at (8mm,-4mm) {$p_3$};
    }} \quad
  \raisebox{-15pt}{\tikz[thick]{
      \draw [left,dashed] (-5mm,5mm)--(0,0);
      \node at (-7mm,4mm) {$p_1$};
      \draw [left,dotted] (-5mm,-5mm)--(0,0);
      \node at (-7mm,-4mm) {$p_2$};
      \draw [right,dotted] (0mm,0)--(5mm,5mm);
      \node at (8mm,4mm) {$p_3$};
      \draw [right,dotted] (0mm,0)--(5mm,-5mm);
      \node at (8mm,-4mm) {$p_4$};
    }}
  \label{dia:source.diagrams}
\end{gather}
We have denoted $A_{rt}$ by dotted and $A_{at}$ by dashed lines. Vertices in the
last line only couple to sources and lead to contact terms in the correlation
functions. The associated Feynmann rules can be obtained from
\eqref{eq:diff-Ls-expanded}.  We will not need the four point interactions in
\eqref{dia:source.diagrams} in the following calculation, but we have enlisted
them anyway for completeness.

We can now utilise \eqref{eq:GreensFunctions} to compute various correlation
functions order-by-order in loops. We are working with the conventions
of~\cite{Crossley:2015evo} for the definition of correlation functions. These
are related to the conventions of~\cite{Wang:1998wg} as
$G^{\text{WH}}_{\alpha\ldots} = i/2\,(-1)^{n_a}(-2i)^{n_r}G_{\alpha\ldots}$ and
those of~\cite{Kovtun:2012rj} as
$G^{\text{K}}_{\alpha\ldots} = (-1)^{n_a}G_{\alpha\ldots}$. We note that the
free propagators $\langle \delta n(p)\varphi_a(-p) \rangle_0$ and
$\langle \delta n(p)\varphi_a(-p) \rangle_0$ in~\cite{Chen-Lin:2018kfl} have
incorrect overall signs compared to our \eqref{eq:diff-props}.  To account for
this, their one-loop results should be modified with
$\lambda\to-\lambda$ and $\lambda'\to-\lambda'$; these are reviewed below.

Note that the ``$ra$'' type propagator
$\langle \delta n(p)\varphi_a(-p) \rangle_0$ in \eqref{eq:diff-props} is purely
retarded while the ``$ar$'' type propagator
$\langle \varphi_a(p)\delta n(-p) \rangle_0$ is purely advanced, which is a
generic feature of these EFTs. This allows us to ignore any diagrams that
contain a loop made entirely of ``$ra$'' or entirely of ``$ar$'' propagators, as
they trivially drop out upon performing the frequency integral with a KMS
consistent renormalisation scheme~\cite{Gao:2018bxz}. Another fact to note is
that the ``$rr$'' propagator $\langle \delta n(p)\delta n(-p) \rangle_0$ can be
decomposed into a retarded and advanced piece as seen in \eqref{eq:diff-props},
which is essentially the statement of FDT~\cite{Wang:1998wg}. This allows us to
drop any diagrams with a single ``$rr$'' propagator closed in a loop, as the
loop integral splits into a purely retarded and a purely advanced piece and
trivially drops out. We shall not enlist such diagrams in our discussion below.


\vspace{1em}

\emph{One-loop 2-point function.}---Let us start the discussion with one-loop
corrections to the retarded 2-point function. At this order there are no
possible diagrams involving a stochastic vertex. However, it is still helpful to
revisit the contribution coming from hydrodynamic diagrams to set up some ground
work (see~\cite{Chen-Lin:2018kfl}). Let us parametrise the loop corrections to
$\langle \delta n(p)\varphi_a(-p) \rangle_0$ in \eqref{eq:diff-props} as
\begin{equation}
  \langle \delta n(p)\varphi_a(-p) \rangle
  \equiv \frac{1}{F(p) + \Sigma(p)}.
  \label{eq:diff2ptAnsatz}
\end{equation}
$\Sigma(p)$ can be understood as a momentum-dependent correction to the
diffusion constant. We have two diagrams that can possibly contribute to this
process
\begin{equation}
  \raisebox{-31pt}{\tikz[thick]{ \draw [right] (2mm,0)--(7mm,0); \draw [aux]
      (7mm,0)--(12mm,0); \node at (7mm,-3mm) {$p$};
      \draw [right] (12mm,0) arc (180:90:6mm); \draw [aux] (18mm,6mm) arc
      (90:0:6mm); \node at (18mm,10mm) {$p'$};
      \draw [right] (12mm,0) arc (180:270:6mm); \draw [aux] (18mm,-6mm) arc
      (-90:0:6mm); \node at (18mm,-9mm) {$p-p'$};
      \draw [right] (24mm,0)--(29mm,0); \draw [aux] (29mm,0)--(34mm,0); \node
      at (29mm,-3mm)
      {$p$}; }} \quad
  \raisebox{-31pt}{\tikz[thick]{ \draw [right] (2mm,0)--(7mm,0); \draw [aux]
      (7mm,0)--(12mm,0); \node at (7mm,-3mm) {$p$};
      \draw [right] (12mm,0) arc (180:90:6mm); \draw [aux] (18mm,6mm) arc
      (90:0:6mm); \node at (18mm,10mm) {$p'$};
      \draw [right] (12mm,0) arc (180:270:6mm); \draw (18mm,-6mm) arc
      (-90:0:6mm); \node at (18mm,-9mm) {$p-p'$};
      \draw [right] (24mm,0)--(29mm,0); \draw [aux] (29mm,0)--(34mm,0); \node
      at (29mm,-3mm)
      {$p$}; }}
  \label{dia:diff-2pt-1loop}
\end{equation}
It is straightforward to compute these and obtain the one-loop correction
\begin{align}
  \Sigma_1(p)
  &= i\lambda T_0 k^2 \int \frac{\df^4p'}{(2\pi)^4}\frac{1}{F(p')} \nn\\
  &\qquad
    \bigg(\frac{\chi \tilde\lambda\, k'\cdot(k-k')}{F(p-p')}
    - \frac{2i\lambda \sigma\, (k-k')^2
    k'^2}{|F(p-p')|^2}
    \bigg) \nn\\
  &\hspace{-2em}
    = i\lambda \chi T_0 k^2 \int \frac{\df^4p'}{(2\pi)^4}
    \frac{\tilde\lambda\, k'\cdot k + (\lambda-\tilde\lambda)k'^2}
    {F(p')F(p-p')} \nn\\
  &\hspace{-2em}
    = \frac{\lambda\chi T_0 k^2 }{32\pi D^2}
    \lb \tilde\lambda (\omega + i D k^2)
    - \lambda\, \omega \rb \sqrt{k^2- \frac{2 i \omega}{D}}.
    \label{eq:Simga1Calc}
\end{align}
In obtaining the second equality, we have expanded the second term in the
brackets into a retarded and advanced piece. The term purely retarded in $p'$
drops out of the integral. The integration has been be performed with a hard
momentum cutoff and cutoff-dependent terms have been ignored; see
\eqref{eq:integrals}.

We can also compute the respective contribution to the retarded two-point
correlation function $G_{ra}$. Using \eqref{eq:GreensFunctions} and
\eqref{eq:diff-Ls-expanded} we can parametrise it as
\begin{align}
  G_{ra}(p)
  &= \frac{-i\delta^2 W}{\delta A_{at}(p)\delta A_{rt}(-p)} \nn\\
  &= -\omega\chi \langle \delta n(p)\varphi_a(-p) \rangle
    + \ldots \nn\\
  &\equiv \frac{ik^2\lb \sigma + \delta\sigma(p)\rb}{F(p) + \Sigma(p)}
    \equiv \frac{ik^2\sigma}{F(p) + \Gamma(p)}.
  \label{eq:2ptFnAnsatz}
\end{align}
Here $\delta\sigma(p)$ is seen as correction to conductivity in the language
of~\cite{Chen-Lin:2018kfl}, while $\Gamma(p)$ is the physically measurable
correction to the two-point function.  The $\ldots$ in the second line
represents contributions involving the source couplings from
\eqref{dia:source.diagrams}, given by diagrams
\begin{equation}
  \begin{gathered}
    \raisebox{-12pt}{\tikz[thick]{
        \draw [right] (-6.1mm,0)--(-6mm,0);
        \draw [dashed] (-6mm,0)--(0,0);
        \node at (-5mm,-3mm) {$p$};
        \draw [dotted,right] (0,0)--(6mm,0);
        \node at (5mm,-3mm) {$p$};
      }}
  \vspace{-1em} \\
    \raisebox{-29pt}{\tikz[thick]{ \draw [right] (2mm,0)--(7mm,0); \draw [aux]
        (7mm,0)--(12mm,0); \node at (7mm,-3mm) {$p$};
        \draw [right] (12mm,0) arc (180:90:6mm); \draw [aux] (18mm,6mm) arc
        (90:0:6mm); \node at (18mm,10mm) {$p'$};
        \draw [right] (12mm,0) arc (180:270:6mm); \draw [aux] (18mm,-6mm) arc
        (-90:0:6mm); \node at (18mm,-9mm) {$p-p'$};
        \draw [right,dotted] (24mm,0)--(29mm,0);
        \node at (29mm,-3mm)
        {$p$};
      }} \quad
    \raisebox{-29pt}{\tikz[thick]{ \draw [right]
        (2mm,0)--(7mm,0); \draw [aux] (7mm,0)--(12mm,0); \node at (7mm,-3mm)
        {$p$};
        \draw [right] (12mm,0) arc (180:90:6mm); \draw [aux] (18mm,6mm) arc
        (90:0:6mm); \node at (18mm,10mm) {$p'$};
        \draw [right] (12mm,0) arc (180:270:6mm); \draw (18mm,-6mm) arc
        (-90:0:6mm); \node at (18mm,-9mm) {$p-p'$};
        \draw [right,dotted] (24mm,0)--(29mm,0);
        \node at (29mm,-3mm) {$p$};
      }}
\end{gathered}
\label{dia:diff-2pt-1loop-coup}
\end{equation}
The first diagram contributes a contact term, while
the remaining two diagrams follows along \eqref{eq:Simga1Calc} leading to
\begin{align}
  (\ldots)
  &= \chi
    + \frac{i\lambda k^2}{F(p)} \int \frac{\df^4p'}{(2\pi)^4} \frac{1}{F(p')}
  \bigg( \frac{-T_0\sigma'k'\cdot(k-k')}{F(p-p')} \nn\\
  &\qquad
  + \frac{2T_0D(k-k')^2 (\chi' \omega'-i\sigma'
    k'\cdot(k-k'))}{|F(p-p')|^2}\bigg) \nn\\
  &= \chi
    - \frac{i\lambda \chi^2 T_0 k^2}{F(p)} \int \frac{\df^4p'}{(2\pi)^4}
    \frac{(\lambda-\tilde\lambda)k'^2}{F(p') F(p-p')}.
\end{align}
In total, we find the correction to the correlation function
\begin{align}
  \Gamma_1(p)
  &= - \frac{\lambda \chi T_0}{D} \int \frac{\df^4p'}{(2\pi)^4}
    \frac{\omega\tilde\lambda\, k'\cdot k
    - iD k^2 (\lambda-\tilde\lambda)k'^2}
    {F(p')F(p-p')} \nn\\
  &= \frac{-\lambda^2\chi T_0}{32\pi D^2} \omega k^2
     \sqrt{k^2- \frac{2 i \omega}{D}}.
    \label{eq:Gamma1Calc}
\end{align}
One can also read out $\delta\sigma$ using \eqref{eq:2ptFnAnsatz}
\begin{align}
  \delta\sigma_1(p)
  &= \frac{\lambda \tilde\lambda \chi^2 T_0}{32\pi D} k^2
    \sqrt{k^2- \frac{2 i \omega}{D}}.
\end{align}
We see that the physical observables exhibit non-analytic behaviour due to
factors of $(k^2-2i\omega/D)^{1/2}$. These are the so called ``long-time tails''
in diffusion model.

These results at $k\neq0$ were originally derived in \cite{Chen-Lin:2018kfl} 
(the long-time tails at $k=0$ is a simpler exercise, see e.g.~\cite{Kovtun:2012rj}).
Instead of using the generating functional, \cite{Chen-Lin:2018kfl} computed the symmetric
function $G_{rr}$ first and used Kramers-Kronig relations and FDT to obtain
$G_{ra}$.

\vspace{1em}

\emph{One-loop 3-point function.}---Moving to the case at point, we want to
probe the signatures of stochastic vertices in hydrodynamic correlation
functions.  The simplest place to look at turns out to be the three-point
``$raa$'' correlator. At tree level, this correlator is controlled by the
hydrodynamic interaction coupling $\lambda$ through
\begin{gather}
  \raisebox{-29pt}{\tikz[thick]{
    \draw [right] (-10mm,0)--(-5mm,0);
    \draw [aux] (-5mm,0)--(0,0);
    \node at (-5mm,-3mm) {$p_1$};
    \draw [right] (0mm,0)--(5mm,5mm);
    \draw [aux] (5mm,5mm)--(10mm,10mm);
    \node at (8mm,4mm) {$p_2$};
    \draw [right] (0mm,0)--(5mm,-5mm);
    \draw [aux] (5mm,-5mm)--(10mm,-10mm);
    \node at (8mm,-4mm) {$p_3$};
  }}
\end{gather}
leading to
\begin{align}
  &\langle \delta n(p_1) \delta\varphi_a(-p_2)\delta\varphi_a(-p_3) \rangle \nn\\
  &\qquad\qquad
    = \frac{-ik_1^2\lb \lambda
    + \delta\lambda(p_1;p_2,p_3)\rb}{F(p_1)F(p_2)F(p_3)},
\end{align}
where $p_2+p_3=p_1$ and $\delta\lambda(p_1;p_2,p_3)$ parametrises possible loop
corrections.  At one loop order, there is only one diagram involving a
stochastic vertex that contributes to this process
\begin{equation}
  \raisebox{-29pt}{\tikz[thick]{
      \draw [right] (-10mm,0)--(-5mm,0);
      \draw [aux] (-5mm,0)--(0,0);
      \node at (-5mm,-3mm) {$p_1$};
      \draw [right] (0,0) arc (180:90:6mm);
      \draw [aux] (6mm,6mm) arc (90:0:6mm);
      \node at (6mm,9mm) {$p'$};
      \draw [right] (0,0) arc (180:270:6mm);
      \draw [aux] (6mm,-6mm) arc (-90:0:6mm);
      \node at (6mm,-8mm) {$p_1-p'$};
      \draw[fill] (12mm,0) circle (.4ex);
      \draw [right] (12mm,0)--(17mm,5mm);
      \draw [aux] (17mm,5mm)--(22mm,10mm);
      \node at (20mm,4mm) {$p_2$};
    \draw [right] (12mm,0)--(17mm,-5mm);
    \draw [aux] (17mm,-5mm)--(22mm,-10mm);
    \node at (20mm,-4mm) {$p_3$};
  }}
\label{eq:diff-3ptdiagram}
\end{equation}
Introducing a hard momentum cutoff, we can compute its contribution to
$\delta\lambda$ as
\begin{align}
  \frac{\delta\lambda_{1}^{\text{st}}}{\lambda}
  &= \frac{2(\vartheta_1+\vartheta_2)}{\chi^2}
    \int \frac{\df^4p'}{(2\pi)^4}
    \frac{(k_2\cdot k_3)\,k'\cdot(k_1-k')}{F(p')F(p_1-p')} \nn\\
  &~
    - \frac{2\vartheta_1}{\chi^2}
    \int \frac{\df^4p'}{(2\pi)^4}
    \frac{(k_1\cdot k_{(2})(k_{3)}\cdot k')
    - (k_2\cdot k')(k_3\cdot k')}{F(p')F(p_1-p')}\nn\\
  &\hspace{-2em}
    = \bigg[
    2 \lb \vartheta_1{+}\vartheta_2 \rb
    (k_2\cdot k_3)\lb k_1^2 - \frac{i\omega_1}{D} \rb
    - \vartheta_1 (k_1\cdot k_{2})(k_1\cdot k_{3}) \nn\\
  &\hspace{-1em}
    - \frac{1}{3}\vartheta_1 (k_2\cdot k_3)
    \lb k_1^2 - \frac{2i\omega_1}{D} \rb
    \bigg] 
    \frac{\sqrt{k_1^2- 2 i \omega_1/D}}{32\pi D\chi^2}.
    \hspace{-0.1em}
\end{align}
Refer \eqref{eq:integrals} for the explicit integral. Again, the cutoff
dependent terms have been ignored. We find a non-analytic correction coming from
the stochastic coefficients $\vartheta_1$, $\vartheta_2$. Note that there can
still be other corrections at this loop order coming from hydrodynamic vertices
that we have not taken into account here.

We can probe the correction $\delta\lambda$ using the three-point retarded
function $G_{raa}$. Using \eqref{eq:GreensFunctions} in conjunction with
\eqref{eq:diff-Ls-expanded}, and keeping track of all the contact terms, we can
find that
\begin{align}
  &G_{raa} (p_1;-p_2,-p_3)
    = \frac{-i\delta^3 W}{\delta A_{at}(p_1)\delta A_{rt}(-p_2)\delta A_{rt}(-p_3)}
    \nn\\
  &\quad
    = \chi^2\omega_2\omega_3
    \langle\delta n(p_1)\varphi_a(-p_2)\varphi_a(-p_3) \rangle
    + \ldots
    \nn\\
  &\quad
    = \frac{-i\chi^2\omega_2\omega_3k_1^2
    \lb \lambda + \delta\lambda(p_1;p_2,p_3)\rb
    }{F(p_1)F(p_2)F(p_3)}
    + \frac{iD k_1^2\chi'}{F(p_1)} \nn\\
  &\qquad
    - \lb\frac{i\omega_2\lb\chi' Dk_1^2 - \sigma' k_1\cdot k_2\rb}{F(p_1)F(p_2)} 
    + (2\leftrightarrow3) \rb.
    \label{G-def-3pt}
\end{align}
Similar to \eqref{eq:2ptFnAnsatz}, the $\ldots$ contributions in the second line
arise due to non-linear background couplings in
\eqref{dia:source.diagrams}. These generate tree-level terms in the last step
due to 
\begin{equation}
  \begin{gathered}
    \raisebox{-19pt}{\tikz[thick]{
      \draw [right] (-5.1mm,0)--(-5mm,0);
      \draw [dashed] (-5mm,0)--(0,0);
      \node at (-5mm,-3mm) {$p_1$};
      \draw [right,dotted] (0mm,0)--(5mm,5mm);
      \node at (8mm,4mm) {$p_2$};
      \draw [right,dotted] (0mm,0)--(5mm,-5mm);
      \node at (8mm,-4mm) {$p_3$};
    }} \qquad
    \raisebox{-29pt}{\tikz[thick]{
        \draw [right] (-5.1mm,0)--(-5mm,0);
        \draw [dashed] (-5mm,0)--(0,0);
        \node at (-5mm,-3mm) {$p_1$};
        \draw [right,dotted] (0mm,0)--(5mm,5mm);
        \node at (8mm,4mm) {$p_2$};
        \draw [right] (0mm,0)--(5mm,-5mm);
        \draw [aux] (5mm,-5mm)--(10mm,-10mm);
        \node at (8mm,-4mm) {$p_3$};
      }} \quad
  \raisebox{-19pt}{\tikz[thick]{
      \draw [right] (-5.1mm,0)--(-5mm,0);
      \draw [dashed] (-5mm,0)--(0,0);
      \node at (-5mm,-3mm) {$p_1$};
      \draw [right] (0mm,0)--(5mm,5mm);
      \draw [aux] (5mm,5mm)--(10mm,10mm);
      \node at (8mm,4mm) {$p_2$};
      \draw [right,dotted] (0mm,0)--(5mm,-5mm);
      \node at (8mm,-4mm) {$p_3$};
    }} \\
  \raisebox{-19pt}{\tikz[thick]{
      \draw [right] (-10mm,0)--(-5mm,0);
      \draw [aux] (-5mm,0)--(0,0);
      \node at (-5mm,-3mm) {$p_1$};
      \draw [right,dotted] (0mm,0)--(5mm,5mm);
      \node at (8mm,4mm) {$p_2$};
      \draw [right,dotted] (0mm,0)--(5mm,-5mm);
      \node at (8mm,-4mm) {$p_3$};
    }} \qquad
    \raisebox{-29pt}{\tikz[thick]{
        \draw [right] (-10mm,0)--(-5mm,0);
        \draw [aux] (-5mm,0)--(0,0);
        \node at (-5mm,-3mm) {$p_1$};
        \draw [right,dotted] (0mm,0)--(5mm,5mm);
        \node at (8mm,4mm) {$p_2$};
        \draw [right] (0mm,0)--(5mm,-5mm);
        \draw [aux] (5mm,-5mm)--(10mm,-10mm);
        \node at (8mm,-4mm) {$p_3$};
      }} \quad
    \raisebox{-19pt}{\tikz[thick]{
        \draw [right] (-10mm,0)--(-5mm,0);
        \draw [aux] (-5mm,0)--(0,0);
        \node at (-5mm,-3mm) {$p_1$};
        \draw [right] (0mm,0)--(5mm,5mm);
        \draw [aux] (5mm,5mm)--(10mm,10mm);
        \node at (8mm,4mm) {$p_2$};
        \draw [right,dotted] (0mm,0)--(5mm,-5mm);
        \node at (8mm,-4mm) {$p_3$};
      }}
  \end{gathered}
\end{equation}
There are, however, no analogous one-loop diagrams involving a stochastic
vertex and background fields.

To get an intuitive feel of this result, we can choose a particular
configuration with $p_1 = 2p_2=2p_3=p$. In the small momentum limit
($k^2 \ll \omega/D$), we get
\begin{align}
  \frac{-2\omega^2}{k^4} \mathrm{Re}\, G_{raa}
  &= \chi^2 D \tilde\lambda
    + \ldots
    + \frac{\lambda
    (\frac{2}{3} \vartheta_1 + \vartheta_2)}{32 \pi D^{5/2}} \omega^{5/2}
    + \ldots, \nn\\
  \frac{4\omega^3}{k^6} \mathrm{Im}\, G_{raa}
  &= \chi^2 D^2 (\lambda +3\tilde{\lambda})
    + \ldots \nn\\
  &\qquad
    + \frac{\lambda (\frac23\vartheta_1+\vartheta_2)}{16 \pi D^{5/2}}
    \frac{\omega^{7/2}}{k^2}
    + \ldots.
\end{align}
as reported in \eqref{eq:intro-G3}. Ellipsis in the middle denote higher
derivative non-stochastic contributions that we have not computed here, while
ellipsis at the end denote terms further suppressed in $k^2$.

\vspace{1em}

\emph{Two-loop 2-point function.}---The quest for stochastic signatures in
2-point functions is considerably more involved. As there are no stochastic
contributions at one-loop order, we need to go to two-loops to get a non-trivial
effect. Focusing on the ``ra'' propagator \eqref{eq:diff2ptAnsatz}, we find 7
independent qualifying diagrams, but only 2 actually contribute. The
non-vanishing diagrams are broadly of three kinds: firstly, we have diagrams
involving a 4-point hydrodynamic interaction
\begin{gather}
  \raisebox{-34pt}{\tikz[thick]{
      \draw [right] (-10mm,0)--(-5mm,0);
      \draw [aux] (-5mm,0)--(0,0);
      \node at (-5mm,-3mm) {$p$};
      \draw [right] (0,0) arc (180:90:7mm);
      \draw [aux] (7mm,7mm) arc (90:0:7mm);
      \node at (8mm,10mm) {$p'$};
      \draw [right] (0mm,0)--(7mm,0);
      \draw [aux] (7mm,0)--(14mm,0);
      \node at (7mm,-3mm) {$p''$};
      \draw [right] (0,0) arc (180:270:7mm);
      \draw [aux] (7mm,-7mm) arc (-90:0:7mm);
      \node at (7mm,-10mm) {$p-p'-p''$};
      \draw[fill] (14mm,0) circle (.4ex);
      \draw [right] (14mm,0)--(19mm,0);
      \draw [aux] (19mm,0)--(24mm,0);
      \node at (19mm,-3mm) {$p$};
    }} \quad
  \raisebox{-34pt}{\tikz[thick]{
      \draw [right] (-10mm,0)--(-5mm,0);
      \draw [aux] (-5mm,0)--(0,0);
      \node at (-5mm,-3mm) {$p$};
      \draw [right] (0,0) arc (180:90:7mm);
      \draw (7mm,7mm) arc (90:0:7mm);
      \node at (8mm,10mm) {$p'$};
      \draw [right] (0mm,0)--(7mm,0);
      \draw [aux] (7mm,0)--(14mm,0);
      \node at (7mm,-3mm) {$p''$};
      \draw [right] (0,0) arc (180:270:7mm);
      \draw [aux] (7mm,-7mm) arc (-90:0:7mm);
      \node at (7mm,-10mm) {$p-p'-p''$};
      \draw[fill] (14mm,0) circle (.4ex);
      \draw [right] (14mm,0)--(19mm,0);
      \draw [aux] (19mm,0)--(24mm,0);
      \node at (19mm,-3mm) {$p$};
    }}
\end{gather}
The only difference between the two diagrams is the stochastic vertex and the
$p'$ propagator. In fact, one can check that due to KMS relations between the
two vertices and the two propagators, the two contributions exactly cancel. Next
we have diagrams involving two 3-point hydrodynamic interactions, where the
stochastic 4-point vertex has an external leg
\begin{gather}
  \raisebox{-29pt}{\tikz[thick]{
      \draw [right] (-14mm,0)--(-9mm,0);
      \draw [aux] (-9mm,0)--(-4mm,0);
      \node at (-9mm,-3mm) {$p$};
      \draw [right] (-4mm,0) arc (180:135:9mm and 7mm);
      \draw [aux] (-1mm,5mm) arc (135:90:7mm);
      \node at (-3mm,7mm) {$p'$};
      \draw [right] (4mm,7mm) arc (120:60:8mm);
      \draw [aux] (16mm,0mm) arc (0:60:8mm);
      \node at (16mm,10mm) {$p'-p''$};
      \draw [right] (4mm,7mm) arc (180:240:8mm);
      \draw [aux] (8mm,0)--(16mm,0);
      \node at (7mm,-3mm) {$p''$};
      \draw [right] (-4mm,0) arc (180:270:10mm and 8mm);
      \draw [aux] (6mm,-8mm) arc (-90:0:10mm and 8mm);
      \node at (6mm,-11mm) {$p-p'$};
      \draw[fill] (16mm,0) circle (.4ex);
      \draw [right] (16mm,0)--(21mm,0);
      \draw [aux] (21mm,0)--(26mm,0);
      \node at (21mm,-3mm) {$p$};
    }} \quad
  \raisebox{-29pt}{\tikz[thick]{
      \draw [right] (-14mm,0)--(-9mm,0);
      \draw [aux] (-9mm,0)--(-4mm,0);
      \node at (-9mm,-3mm) {$p$};
      \draw [right] (-4mm,0) arc (180:135:9mm and 7mm);
      \draw [aux] (-1mm,5mm) arc (135:90:7mm);
      \node at (-3mm,7mm) {$p'$};
      \draw [right] (4mm,7mm) arc (120:60:8mm);
      \draw [aux] (16mm,0mm) arc (0:60:8mm);
      \node at (16mm,10mm) {$p'-p''$};
      \draw [right] (4mm,7mm) arc (180:240:8mm);
      \draw [aux] (8mm,0)--(16mm,0);
      \node at (7mm,-3mm) {$p''$};
      \draw [right] (-4mm,0) arc (180:270:10mm and 8mm);
      \draw (6mm,-8mm) arc (-90:0:10mm and 8mm);
      \node at (6mm,-11mm) {$p-p'$};
      \draw[fill] (16mm,0) circle (.4ex);
      \draw [right] (16mm,0)--(21mm,0);
      \draw [aux] (21mm,0)--(26mm,0);
      \node at (21mm,-3mm) {$p$};
    }}
  \nn\\
  \raisebox{-29pt}{\tikz[thick]{
      \draw [right] (-14mm,0)--(-9mm,0);
      \draw [aux] (-9mm,0)--(-4mm,0);
      \node at (-9mm,-3mm) {$p$};
      \draw [right] (-4mm,0) arc (180:135:9mm and 7mm);
      \draw [aux] (-1mm,5mm) arc (135:90:7mm);
      \node at (-3mm,7mm) {$p'$};
      \draw [right] (4mm,7mm) arc (120:60:8mm);
      \draw (16mm,0mm) arc (0:60:8mm);
      \node at (16mm,10mm) {$p'-p''$};
      \draw [right] (4mm,7mm) arc (180:240:8mm);
      \draw [aux] (8mm,0)--(16mm,0);
      \node at (7mm,-3mm) {$p''$};
      \draw [right] (-4mm,0) arc (180:270:10mm and 8mm);
      \draw [aux] (6mm,-8mm) arc (-90:0:10mm and 8mm);
      \node at (6mm,-11mm) {$p-p'$};
      \draw[fill] (16mm,0) circle (.4ex);
      \draw [right] (16mm,0)--(21mm,0);
      \draw [aux] (21mm,0)--(26mm,0);
      \node at (21mm,-3mm) {$p$};
    }}
\end{gather}
These diagrams also only differ from each other in the stochastic vertex
involved and the propagators arriving at it, and mutually cancel due to KMS
properties. Finally we have diagrams involving two 3-point hydrodynamic vertices
but no external leg on the stochastic vertex
\begin{equation}
\begin{gathered}
  \raisebox{-29pt}{\tikz[thick]{
      \draw [right] (-10mm,0)--(-5mm,0);
      \draw [aux] (-5mm,0)--(0,0);
      \node at (-5mm,-3mm) {$p$};
      \draw [right] (0,0) arc (180:90:6mm);
      \draw [aux] (6mm,6mm) arc (90:0:6mm);
      \node at (6mm,10mm) {$p'$};
      \draw [right] (0,0) arc (180:270:6mm);
      \draw [aux] (6mm,-6mm) arc (-90:0:6mm);
      \node at (6mm,-9mm) {$p-p'$};
      \draw[fill] (12mm,0) circle (.4ex);
      \draw [right] (12mm,0) arc (180:90:6mm);
      \draw [aux] (18mm,6mm) arc (90:0:6mm);
      \node at (18mm,10mm) {$p''$};
      \draw [right] (12mm,0) arc (180:270:6mm);
      \draw [aux] (18mm,-6mm) arc (-90:0:6mm);
      \node at (18mm,-9mm) {$p-p''$};
      \draw [right] (24mm,0)--(29mm,0);
      \draw [aux] (29mm,0)--(34mm,0);
      \node at (29mm,-3mm) {$p$};
    }} \\
  \raisebox{-29pt}{\tikz[thick]{
      \draw [right] (-10mm,0)--(-5mm,0);
      \draw [aux] (-5mm,0)--(0,0);
      \node at (-5mm,-3mm) {$p$};
      \draw [right] (0,0) arc (180:90:6mm);
      \draw [aux] (6mm,6mm) arc (90:0:6mm);
      \node at (6mm,10mm) {$p'$};
      \draw [right] (0,0) arc (180:270:6mm);
      \draw [aux] (6mm,-6mm) arc (-90:0:6mm);
      \node at (6mm,-9mm) {$p-p'$};
      \draw[fill] (12mm,0) circle (.4ex);
      \draw [right] (12mm,0) arc (180:90:6mm);
      \draw [aux] (18mm,6mm) arc (90:0:6mm);
      \node at (18mm,10mm) {$p''$};
      \draw [right] (12mm,0) arc (180:270:6mm);
      \draw (18mm,-6mm) arc (-90:0:6mm);
      \node at (18mm,-9mm) {$p-p''$};
      \draw [right] (24mm,0)--(29mm,0);
      \draw [aux] (29mm,0)--(34mm,0);
      \node at (29mm,-3mm) {$p$};
    }}
\end{gathered}
\end{equation}
These diagrams involve different hydrodynamic vertices, which each come with an
independent coefficient, and hence their contribution to the propagator does not
cancel. In fact we have already computed parts of these diagrams. The left half
including the stochastic vertex is just the 3-point diagram
\eqref{eq:diff-3ptdiagram}. Taking it into account, these diagrams are just the
one-loop diagrams in \eqref{dia:diff-2pt-1loop} with $\lambda$ in the left
three-point vertex replaced by $\delta\lambda$ before performing the
integral. Hence, we find the stochastic correction to the propagator using
\eqref{eq:Simga1Calc} as
\begin{align}
  \Sigma_2^{\text{st}}(p)
  &= iT_0\chi k^2 \int \frac{\df^4p'}{(2\pi)^4}
    \delta\lambda_1^{\text{st}}(p;p',p-p') \nn\\
  &\qquad\qquad
    \frac{\tilde\lambda\, k'\cdot k + (\lambda-\tilde\lambda) k'^2}
    {\Delta(p')\Delta(p-p')} \nn\\
  &= \frac{T_0 \lambda  k^2}{1024 \pi^2 D^3\chi}
    \lb (\lambda-\tilde\lambda) \omega -i\tilde\lambda D k^2 \rb
    \lb k^2 - \frac{2i\omega}{D} \rb
     \nn\\
  &\quad
    \lb \frac16 \vartheta_1 k^4 - \lb \frac23 \vartheta_1+\vartheta_2 \rb
    \lb k^2 - \frac{i\omega}{D} \rb^2 
    \rb.
\end{align}

Similarly, the stochastic correction to the retarded function $G_{ra}$ involves
background coupling diagrams
\begin{equation}
  \begin{gathered}
      \raisebox{-29pt}{\tikz[thick]{
      \draw [right] (-10mm,0)--(-5mm,0);
      \draw [aux] (-5mm,0)--(0,0);
      \node at (-5mm,-3mm) {$p$};
      \draw [right] (0,0) arc (180:90:6mm);
      \draw [aux] (6mm,6mm) arc (90:0:6mm);
      \node at (6mm,10mm) {$p'$};
      \draw [right] (0,0) arc (180:270:6mm);
      \draw [aux] (6mm,-6mm) arc (-90:0:6mm);
      \node at (6mm,-9mm) {$p-p'$};
      \draw[fill] (12mm,0) circle (.4ex);
      \draw [right] (12mm,0) arc (180:90:6mm);
      \draw [aux] (18mm,6mm) arc (90:0:6mm);
      \node at (18mm,10mm) {$p''$};
      \draw [right] (12mm,0) arc (180:270:6mm);
      \draw [aux] (18mm,-6mm) arc (-90:0:6mm);
      \node at (18mm,-9mm) {$p-p''$};
      \draw [right,dotted] (24mm,0)--(29mm,0);
      \node at (29mm,-3mm) {$p$};
    }} \\
  \raisebox{-29pt}{\tikz[thick]{
      \draw [right] (-10mm,0)--(-5mm,0);
      \draw [aux] (-5mm,0)--(0,0);
      \node at (-5mm,-3mm) {$p$};
      \draw [right] (0,0) arc (180:90:6mm);
      \draw [aux] (6mm,6mm) arc (90:0:6mm);
      \node at (6mm,10mm) {$p'$};
      \draw [right] (0,0) arc (180:270:6mm);
      \draw [aux] (6mm,-6mm) arc (-90:0:6mm);
      \node at (6mm,-9mm) {$p-p'$};
      \draw[fill] (12mm,0) circle (.4ex);
      \draw [right] (12mm,0) arc (180:90:6mm);
      \draw [aux] (18mm,6mm) arc (90:0:6mm);
      \node at (18mm,10mm) {$p''$};
      \draw [right] (12mm,0) arc (180:270:6mm);
      \draw (18mm,-6mm) arc (-90:0:6mm);
      \node at (18mm,-9mm) {$p-p''$};
      \draw [right,dotted] (24mm,0)--(29mm,0);
      \node at (29mm,-3mm) {$p$};
    }}
  \end{gathered}
\end{equation}
Their contribution, again, is just given by substituting $\lambda$ in the left
vertex in \eqref{dia:diff-2pt-1loop-coup} with $\delta\lambda$. We can,
therefore, find stochastic correction to the retarded correlation function using
\eqref{eq:Gamma1Calc} as
\begin{align}
  \Gamma_2^{\text{st}}(p)
  &= \frac{- \chi T_0}{D} \int \frac{\df^4p'}{(2\pi)^4}
    \delta\lambda_1^{\text{st}}(p;p',p-p') \nn\\
  &\qquad\qquad
    \frac{\omega\tilde\lambda\, k'\cdot k
    - iD k^2 (\lambda-\tilde\lambda)k'^2}
    {F(p')F(p-p')} \nn\\
  &= \frac{T_0 \lambda^2 \omega k^2}{1024\pi^2 D^3\chi}
    \lb k^2 - \frac{2i\omega}{D} \rb
     \nn\\
  &\quad
    \lb \frac16 \vartheta_1 k^4 - \lb \frac23\vartheta_1+\vartheta_2 \rb
    \lb k^2 - \frac{i\omega}{D} \rb^2 
    \rb.
\end{align}
One can also find the associated $\delta\sigma$ using \eqref{eq:2ptFnAnsatz} to
be
\begin{align}
  \delta\sigma_2^{\text{st}}(p)
  &= - \frac{T_0 \lambda \tilde\lambda k^2}{1024 \pi^2 D^2}
    \lb k^2 - \frac{2i\omega}{D} \rb
     \nn\\
  &~~
    \lb \frac16 \vartheta_1 k^4 - \lb \frac23 \vartheta_1+\vartheta_2 \rb
    \lb k^2 - \frac{i\omega}{D} \rb^2 
    \rb.
\end{align}
We see that the two-point function gets an analytic correction due to stochastic
vertices. Schematically, the two loops individually have a square-root
non-analyticity, but they combine to give an analytic correction. Accordingly,
it is expected that non-analytic stochastic contributions will kick in the
two-point function at three-loop order.

We can combine these expressions with the one-loop results and expand in
$k^2 \ll \omega/D$ limit. Through a straight-forward computation, we find
\begin{align}
  \frac{\omega^2}{k^4} \mathrm{Re}\,G_{ra}
  &= \chi D^2
    + \frac{\chi^2\lambda^2 T_0}{32 \pi D^{3/2}} \omega^{3/2}
    + \ldots \nn\\
  &\qquad
    - \frac{\lambda ^2 T_0 (\frac23\vartheta_1+\vartheta_2)}{512 \pi^2 D^5}
    \omega^4
    + \ldots, \nn\\
  \frac{\omega}{k^2} \mathrm{Im}\,G_{ra}
  &= \chi D
    + \frac{\chi^2\lambda^2 T_0}{32 \pi D^{3/2}} \omega^{1/2}k^2
    + \ldots \nn\\
  &\qquad
    -\frac{\lambda^2 T_0 (\frac23\vartheta_1+\vartheta_2)}{1024 \pi ^2 d^4}
    \omega^2 k^4
    + \ldots,
\end{align}
reproducing \eqref{eq:intro-G2}. Middle ellipsis denote further non-stochastic
corrections to the correlator that we have not considered here.

\vspace{1em}

\emph{Tree-level 4-point functions.}---Let us finally look at the stochastic
contributions to tree-level 4-point functions. It is clear that stochastic
vertices do not contribute to retarded functions at tree-level. However they do
contribute to non-retarded ones. For instance, we can compute the contribution
to the partially-retarded 4-point function $G_{rraa}$ through the diagram
\begin{equation}
  \raisebox{-26pt}{\tikz[thick]{
      \draw [right] (-10mm,10mm)--(-5mm,5mm);
      \draw [aux] (-5mm,5mm)--(0,0);
      \node at (-8mm,4mm) {$p_1$};
      \draw [right] (-10mm,-10mm)--(-5mm,-5mm);
      \draw [aux] (-5mm,-5mm)--(0,0);
      \node at (-8mm,-4mm) {$p_2$};
      \draw [aux] (10mm,10mm)--(5mm,5mm);
      \draw [right] (0mm,0)--(5mm,5mm);
      \node at (8mm,4mm) {$p_3$};
      \draw [aux] (10mm,-10mm)--(5mm,-5mm);
      \draw [right] (0mm,0)--(5mm,-5mm);
      \node at (8mm,-4mm) {$p_4$};
      \draw[fill] (0,0) circle (.4ex);
    }}
\end{equation}
This evaluates to
\begin{align}
  G_{rraa}&(p_1,p_2;-p_3,-p_4) \nn\\
  &= \frac{-\delta^4 W}{\delta A_{at}(p_1)\delta A_{at}(p_2)
    \delta A_{rt}(-p_3)\delta A_{rt}(-p_4)} \nn\\
  &= \chi^2 \omega_3\omega_4 \langle \delta n(p_1) \delta n(p_2)
    \varphi_a(-p_3)\varphi_a(-p_4)\rangle \nn\\
  &= \ldots + \frac{4(\vartheta_1+\vartheta_2)\omega_3\omega_4
    (k_1\cdot k_2)(k_3\cdot k_4)
    }{F(p_1)F(p_2)F(p_3)F(p_4)} \nn\\
  &\qquad
    - \frac{4\vartheta_1
    \omega_3\omega_4 (k_1\cdot k_{(3})(k_{4)}\cdot k_2) 
    }{F(p_1)F(p_2)F(p_3)F(p_4)}.
    \label{eq:Grraa}
\end{align}
Note that there are no possible contact terms due to \eqref{dia:source.diagrams}
in this computation.  Ellipsis denote that we have ignored any contribution
coming from non-stochastic interactions. Let us take $p_1=p_4 = (\omega, k_1)$
and $p_2=p_3 = (\omega, k_2)$ with $|k_1|=|k_2| = k$ and
$k_1\cdot k_2 = k^2\cos\theta$, corresponding to two propagating density packets
intersecting each other at angle $\theta$. This reduces to
\begin{equation}
  G_{rraa}
  = \ldots + \frac{2\omega^2 k^4}{(\omega+iDk^2)^4}
    \lb 2\vartheta_2 \cos^2\theta
    - \vartheta_1 \sin^2\theta \rb.
\end{equation}
There can be other contributions coming from non-stochastic hydrodynamic
vertices as well.


\vspace{1em}

\emph{Integrals.}---The following loop integrals were used in the calculation
above. The integrals have been performed with a hard momentum cutoff and
cutoff-dependent terms have been ignored. We have
\begin{subequations}
  \begin{align}
    \int\frac{\df^4p'}{(2\pi)^4} \frac{k'^i}{F(p')F(p-p')} 
    &= \frac{k^i}{32\pi D} \sqrt{k^2- \frac{2 i \omega}{D}}, \\
    \int\frac{\df^4p'}{(2\pi)^4} \frac{k'^i k'^j}{F(p')F(p-p')}
    &= \frac{1}{64\pi D} \sqrt{k^2- \frac{2 i \omega}{D}} \nn\\
    &\hspace{-2em}
      \lB k^i k^j - \frac{1}{3} k^2 \delta^{ij}
      + \delta^{ij}\frac{2 i \omega}{3D}\rB, \\
    \int\frac{\df^4p'}{(2\pi)^4} \frac{k'^i k'^j k'^k}{F(p')F(p-p')}
    &= \frac{1}{128\pi D} \sqrt{k^2- \frac{2 i \omega}{D}}
      \nn\\
    &\hspace{-5em}
      \lB k^i k^j k^k
      - \lb k^2 - \frac{2i\omega}{D} \rb k^{(i}\delta^{jk)} \rB, \\
    \int\frac{\df^4p'}{(2\pi)^4} \frac{k'^i k'^j k'^k k'^l}{F(p')F(p-p')}
    &= \frac{1}{256\pi D} \sqrt{k^2- \frac{2 i \omega}{D}} \nn\\
    &\hspace{-4em}
       \bigg[
      k^i k^j k^k k^l
      -2 \lb k^2-\frac{2i\omega }{D} \rb k^{(i}k^j\delta^{kl)} \nn\\
    &\hspace{-2em}
      + \frac15 \left(k^2 - \frac{2 i \omega}{D} \right)^2
      \delta^{(ij}\delta^{kl)} \bigg].
  \end{align}
  \label{eq:integrals}%
\end{subequations}%
Here $p = (\omega,k)$, $p' = (\omega',k')$, and $F(p) = \omega + iDk^2$. Round
brackets denote symmetrisation over all the enclosing indices, divided by the
number of permutations.



\section{Details of KMS block manipulations}

In this appendix we elaborate on some calculational details regarding the
derivation of KMS blocks.

We note that the most general Lagrangian consistent with \eqref{eq:SK1} can be
arranged in a power series in $\Phi_a$ starting from the linear term with
appropriate powers of $i$ to account for the imaginary part
\begin{equation}
  \mathcal{L} = \sum_{m=1}^\infty (-i)^{m+1}
  \mathcal{F}_m(\underbrace{\Phi_a,\Phi_a,\ldots}_{\times m}).
  \label{eq:L-Fm}
\end{equation}
Here $\cF_m$ are a set of totally symmetric real multi-linear maps, with
allowing $m$ arguments, made out of $\Phi_r$ and $\beta^\mu$. The
``under-brace'' notation represents $m$ identical arguments. From here, one can
check that \eqref{eq:L-G} follows by a resummation of \eqref{eq:L-Fm} with 
\begin{equation}
  \cF_{m}(\underbrace{\Phi_a,\ldots}_{\times m})
  = \sum_{n=m}^{2m+1} c_{mn}
  \cD_{n}(\underbrace{\Phi_a,\ldots}_{\times m},
  \underbrace{\lie_\beta\Phi_r,\ldots}_{\times n-m}).
\end{equation}
where $c_{mn}$ is an invertible matrix given by 
\begin{equation}
  c_{mn} = 
  \begin{cases}
    (-1)^{n/2+1} {\textstyle\binom{n/2}{n-m}}
    & \text{for $n$ even}, \\
    (-1)^{(n+1)/2} \frac{m+1}{n+1} {\textstyle\binom{(n+1)/2}{n-m}}
    & \text{for $n$ odd}.
  \end{cases}
\end{equation}
$\lceil n/2\rceil$ and $\lfloor n/2\rfloor$ denote ceiling and floor functions.

To illustrate the second law of thermodynamics, we note that we can use
integration by parts to define
\begin{align}
  \mathcal{D}_n(\Phi_a,\ldots)
  &= \half\mathcal{T}^{\mu\nu}_n(\ldots) G_{a\mu\nu}
  + \mathcal{J}^\mu_n(\ldots) B_{a\mu}  \nn\\
  &\qquad
    + \nabla_\mu\mathcal{N}_n^\mu(\Phi_a,\ldots).
    \label{eq:byparts}
\end{align}
This equation essentially says that if $\mathcal{D}_n$ acts on $\Phi_a$ as a
derivative operator, we can always extract out $\Phi_a$ by adding total
derivatives. This allows us to derive the classical constitutive relations
performing Euler-Lagrange derivatives of \eqref{eq:L-G} and setting ``$a$'' type
fields to zero
\begin{align}
  T^{\mu\nu}
  &= \cT^{\mu\nu}_1 - \cT^{\mu\nu}_{2}(\lie_\beta\Phi_r) 
    - \frac12 \cT^{\mu\nu}_{3}(\lie_\beta\Phi_r,\lie_\beta\Phi_r), \nn\\
  J^{\mu}
  &= \cJ^{\mu}_1 - \cJ^{\mu}_{2}(\lie_\beta\Phi_r) 
    - \frac12 \cJ^{\mu}_{3}(\lie_\beta\Phi_r,\lie_\beta\Phi_r).
    \label{eq:classical-consti}
\end{align}
Note that operators higher than $n=3$ do not appear in the constitutive
relations. In terms of these, the entropy current is defined as
\begin{align}
  S^\mu
  &= - \frac{1}{T} \lb T^{\mu\nu}u_\nu + \mu J^\mu \rb
  + \mathcal{N}^\mu_0
  + \mathcal{N}^\mu_1(\lie_\beta \Phi_r) \nn\\
  &\hspace{-2em}
  - \mathcal{N}^\mu_2(\lie_\beta \Phi_r,\lie_\beta \Phi_r)
  - \half \mathcal{N}^\sM_3(\lie_\beta \Phi_r,\lie_\beta \Phi_r,
    \lie_\beta \Phi_r).
    \hspace{-0.1em}
\end{align}
We can reuse \cref{eq:byparts}, with $\Phi_a$ replaced by $\lie_\beta\Phi_r$,
and the classical conservation equations
$\nabla_\mu T^{\mu\nu} = F^{\nu\rho}J_\rho$, $\nabla_\mu J^\mu = 0$ to derive
the second law statement in \eqref{eq:EC}.

To derive the KMS blocks in \eqref{eq:L1-general} and \eqref{eq:L2-general}, we
expand $\mathcal{D}_n$ into components
\begin{equation}
  \cD_{n}(\underbrace{\circ,\ldots}_{\times n})
  = \sum_{m=0}^{n} \cD_{n,m}(\underbrace{\circ,\ldots}_{\times n}).
\end{equation}
Here ``$\circ$'' is a placeholder for an arbitrary argument of the operator. As
mentioned in the main text, $\cD_{n,m}$ can be seen as the component of
$\mathcal{D}_n$ which has $m$ of its arguments projected transverse to
$\lie_\beta\Phi_r$. We can insert this into \eqref{eq:L-G} and obtain
\begin{align}
  \mathcal{L}
  &= \cD_{1,1}
    + \sum_{n=1}^\infty \sum_{m=0}^{2n} i\cD_{2n,m} 
    + \sum_{n=1}^\infty \sum_{m=0}^{2n+1} \cD_{2n+1,m}.
    \label{eq:L-components}
\end{align}
Note that $\cD_{1,0} = 0$. The arguments of $\mathcal{D}_{n,m}$ are same as
$\mathcal{D}_n$ in \eqref{eq:L-G} for all $m$. Let us first consider the
$\mathcal{D}_{2n,m}$ contribution and use the multi-linear nature of $\mathcal{D}_{n}$ to expand as
\begin{align}
  \mathcal{D}_{2n,m}
  &(\underbrace{\Phi_a,\ldots}_{\times n},
    \underbrace{\Phi_a {+} i \lie_\beta\Phi_r,
    \ldots}_{\times n}) \nn\\
  &= \sum_{s=0}^{n} \sbinom{n}{s}
    \mathcal{D}_{2n,m}(\underbrace{\Phi_a,\ldots}_{\times 2n-s},
    \underbrace{i\lie_\beta\Phi_r,
    \ldots}_{\times s}).
\end{align}
The summation maxes out at $s=2n-m$ as $\mathcal{D}_{2n,m}$ vanishes for higher
number of $\lie_\beta\Phi_r$ in its arguments. For $m\leq n$ the summation never
reaches this limit and $\mathcal{D}_{2n,m}$ for $m\leq n$ falls in the $n$th KMS
block. $\mathcal{D}_{2n,m}$ for $m>n$, however, has the least power of $\Phi_a$
to be $m$ and falls in the $m$th KMS block.  Similarly for
$\mathcal{D}_{2n+1,m}$ ($n\neq 0$), we find
\begin{align}
  &\cD_{2n+1,m}
  (\Phi_a{+}{\textstyle\frac{i}{2}}\lie_\beta\Phi_r,
    \underbrace{\Phi_a,\ldots}_{\times n},
    \underbrace{\Phi_a {+} i \lie_\beta\Phi_r,\ldots}_{\times n}) \nn\\
  &\quad
    = \sum_{s=0}^{n} \sbinom{n}{s} \cD_{2n+1,m}
    (\underbrace{\Phi_a,\ldots}_{\times 2n+1-s},
    \underbrace{i\lie_\beta\Phi_r,\ldots}_{\times s}) \nn\\
  &~~\quad + \half \sum_{s=1}^{n+1} \sbinom{n}{s-1} \cD_{2n+1,m}
    (\underbrace{\Phi_a,\ldots}_{\times 2n+1-s},
    \underbrace{i\lie_\beta\Phi_r,\ldots}_{\times s}).
\end{align}
In this case, the summations max out at $s=2n+1-m$. For $m\leq n$, we never hit
this limit and hence $\mathcal{D}_{2n+1,m}$ for $m\leq n$ falls in the $n$th KMS
block. For $m>n$, however, we find that $\mathcal{D}_{2n+1,m}$ falls in the
$m$th KMS block. The same result also applies to $\mathcal{D}_{1,1}$ which falls
in the 1st KMS block. We can now rearrange \eqref{eq:L-components} according to
the contributions to KMS blocks and obtain
\begin{align}
  \mathcal{L}
  &= \sum_{n=1}^\infty \Bigg[
    \sum_{m=\lceil n/2\rceil}^{n-1} i\cD_{2m,n} +
    \sum_{\mathclap{m=\lfloor n/2\rfloor}}^{n-1} \cD_{2m+1,n} \nn\\
  &\qquad\qquad
    + \sum_{m=0}^{n}
    \lb i\cD_{2n,m} {+} \cD_{2n+1,m} \rb
    \Bigg].
    \label{eq:KMS-resummation}
\end{align}

It is tempting to identify the expression in the square brackets in
\eqref{eq:KMS-resummation} as the $n$th KMS block. However, there is a final
subtlety we need to account for. Consider, for instance, the first term in the
series, with the arguments from \cref{eq:L-G} put back in
\begin{align}
  &\cD_{1,1}(\Phi_a)
    + i\cD_{2,0}(\Phi_a,\Phi_a{+}i\lie_\beta\Phi_r)
    + i\cD_{2,1}(\Phi_a,\Phi_a{+}i\lie_\beta\Phi_r)\nn\\
  &\quad
     + \cD_{3,0}(\Phi_a{+}{\textstyle\frac{i}{2}}\lie_\beta\Phi_r,
    \Phi_a,\Phi_a{+}i\lie_\beta\Phi_r) \nn\\
  &\quad
    + \cD_{3,1}(\Phi_a{+}{\textstyle\frac{i}{2}}\lie_\beta\Phi_r,
    \Phi_a,\Phi_a{+}i\lie_\beta\Phi_r).
    \label{eq:L1-raw}
\end{align}
We can push the $\mathcal{D}_{3,1}$ term to $\mathcal{L}_2$ by redefining
$\mathcal{D}_{1,1}$
\begin{equation}
  \cD_{1,1}(\circ)
  \to \cD_{1,1}(\circ)
  + \half \cD_{3,1}(\circ,\lie_\beta\Phi_r,\lie_\beta\Phi_r).
\end{equation}
This is still a $\Theta$-even linear operator that vanishes upon replacing its
argument with $\lie_\beta\Phi_r$, as required by the KMS structure. Note that we
cannot push $\mathcal{D}_{2,1}$ to $\mathcal{L}_2$ in a similar manner because
that would require shifting $\cD_{1,1}$ with a $\Theta$-odd term. The remaining
terms in \eqref{eq:L1-raw} make up the first KMS block given in
\eqref{eq:L1-general}. Similarly, we can work out the second term in the series
in \eqref{eq:KMS-resummation}, including $\mathcal{D}_{3,1}$ shifted from the
previous term, and obtain
\begin{align}
  &i\cD_{2,2}(\Phi_a,\Phi_a)
    + \cD_{3,2}(\Phi_a,\Phi_a,\Phi_a{+}{\textstyle\frac{3i}{2}}\lie_\beta\Phi_r)
    \nn\\
  &+ \sum_{m=0}^{2}
    i\cD_{4,m}(\Phi_a,\Phi_a,\Phi_a{+}i\lie_\beta\Phi_r,
    \Phi_a{+}i\lie_\beta\Phi_r)  \nn\\
  &+ \sum_{m=0}^{2}\cD_{5,m}(\Phi_a{+}{\textstyle\frac{i}{2}}\lie_\beta\Phi_r,
    \Phi_a,\Phi_a,\Phi_a{+}i\lie_\beta\Phi_r,
    \Phi_a{+}i\lie_\beta\Phi_r) \nn\\
  &+ \cD_{3,1}(\Phi_a,\Phi_a,\Phi_a{+}{\textstyle\frac{3i}{2}}\lie_\beta\Phi_r).
\end{align}
We can again shift
\begin{gather}
  \mathcal{D}_{2,2}(\circ,\circ)
  = \mathcal{D}_{2,2}(\circ,\circ)
  + \cD_{4,2}(\circ,\circ,\lie_\beta\Phi_r,\lie_\beta\Phi_r), \nn\\
  \mathcal{D}_{3,1}(\circ,\circ,\circ)
  = \mathcal{D}_{3,1}(\circ,\circ,\circ)
  + \frac13\cD_{5,1}(\circ,\circ,\circ,\lie_\beta\Phi_r,\lie_\beta\Phi_r), \nn\\
  \mathcal{D}_{3,2}(\circ,\circ,\circ)
  = \mathcal{D}_{3,2}(\circ,\circ,\circ)
  + \frac13\cD_{5,2}(\circ,\circ,\circ,\lie_\beta\Phi_r,\lie_\beta\Phi_r),
\end{gather}
to push $\mathcal{D}_{4,2}$, $\mathcal{D}_{5,1}$, and $\mathcal{D}_{5,2}$ to
$\mathcal{L}_3$ and obtain the full expression for $\mathcal{L}_2$ as
\begin{align}
  \mathcal{L}_2
  &= i\cD_{2,2}(\Phi_a,\Phi_a)
    + \cD_{3,2}(\Phi_a,\Phi_a,\Phi_a{+}{\textstyle\frac{3i}{2}}\lie_\beta\Phi_r)
    \nn\\
  &+ i\cD_{4,0}(\Phi_a,\Phi_a,\Phi_a{+}i\lie_\beta\Phi_r,
    \Phi_a{+}i\lie_\beta\Phi_r)  \nn\\
  &+ i\cD_{4,1}(\Phi_a,\Phi_a,\Phi_a{+}i\lie_\beta\Phi_r,
    \Phi_a{+}i\lie_\beta\Phi_r)  \nn\\
  &+ \cD_{5,0}(\Phi_a{+}{\textstyle\frac{i}{2}}\lie_\beta\Phi_r,
    \Phi_a,\Phi_a,\Phi_a{+}i\lie_\beta\Phi_r,
    \Phi_a{+}i\lie_\beta\Phi_r) \nn\\
  &+ \cD_{3,1}(\Phi_a,\Phi_a,\Phi_a{+}{\textstyle\frac{3i}{2}}\lie_\beta\Phi_r).
\end{align}
The expression for $\mathcal{L}_2$ given in \eqref{eq:L2-general} in the main
text has been truncated to the leading derivative order. This procedure can be
in principle be iterated to obtain higher KMS blocks as well.


\clearpage

\end{document}